\documentclass[aps,pre,twocolumn,10pt,nofootinbib,superscriptaddress,floatfix]{revtex4-2}
\usepackage{latexsym}
\usepackage{amsmath}
\usepackage{amsfonts}

\usepackage{hhline}
\usepackage{graphicx}
\usepackage{amssymb}
\usepackage{bm}
\usepackage{xr-hyper}
\usepackage{color}

\begin{document}
\title{Entropic tug of war: Topological constraints spontaneously rectify the dynamics of a polymer with heterogeneous fluctuations.}

\author{Adam, H. T. P. H\"{o}fler}
\affiliation{Faculty of Physics, University of Vienna, Boltzmanngasse 5, A-1090 Vienna, Austria}
\author{Iurii Chubak}
\affiliation{Faculty of Physics, University of Vienna, Boltzmanngasse 5, A-1090 Vienna, Austria}
\author{Christos N. Likos}
\affiliation{Faculty of Physics, University of Vienna, Boltzmanngasse 5, A-1090 Vienna, Austria}
\author{Jan Smrek}
\email{jan.smrek@univie.ac.at}
\affiliation{Faculty of Physics, University of Vienna, Boltzmanngasse 5, A-1090 Vienna, Austria}


\begin{abstract}
Polymers with active segments constitute prospective future materials and are used as a model for some biological systems such as chromatin. The directions of the active forces are typically introduced with temporal or spatial correlations to establish directional motion of the chain and corresponding active dynamics. Instead, here we consider an active-passive copolymer, where the two segments differ only by the magnitude of their fluctuations and feature no artificial  correlations. Here we show that although the model itself does not possess directional dynamics, if the chains are concentrated, directional persistent motion spontaneously arises as a consequence of the broken translational symmetry owing to the topological constraints. Using scaling arguments and simulations, we explain the phenomenon and describe the ensuing dynamics. Our work has thus far-reaching consequences for the mechanical properties of all dense active polymeric systems with heterogeneous fluctuations and in particular for chromatin conformation and dynamics that are crucial for biological functionality.
\end{abstract}

\maketitle

Active polymers, encountered in nature as chromatin fiber, cytoskeleton, or living worms \cite{EshghiPRL23,Fuerthauer2019, Patil_Science23}, form a subclass of active matter characterized by out-of-equilibrium driving on the single filament level, sharing many of the collective phenomena found in colloidal active matter \cite{Ramaswamy,Active_Matter_review_RevModPhys13}.  
Besides biology, activity seems a promising control parameter for the tuning of the constituent material properties. In particular, the polymers feature topological constraints, stemming from the chains' uncrossability that, as a result, in passive systems govern the stress relaxation and their viscoelastic properties \cite{WinklerGompper2017}. 
While the active, \emph{stiff} filaments the uncrossability in mutual collisions generates torques and ordering, the \emph{flexible} polymers that we focus on in here allow for multi-chain entanglements affecting the material properties in various contrasting ways. The energy input owing to the activity typically speeds up the (entanglement) relaxation, due to the active propagation of the chain along its contour, thereby decreasing the apparent viscosity \cite{Active_Fluidization_Humphrey_NAT02,Suvendu,Tejedor_Ramirez_Pol23}. In contrast, the activity can create new forms of entanglements, inaccessible in equilibrium, generating novel phases of matter, such as the active topological glass \cite{ATG_NatComm20,ATG_PRR20,Micheletti_ACSML24}. 

The directional and persistent motion of the chains lying behind both contrasting trends in material properties is typically introduced in the microscopic model by hand, e.g., by imposing the active force/velocity to act in the backbone direction \cite{Locatelli2018,Lamura2024,Ubertini_ACSML24,Suvendu,Tejedor_Ramirez_Pol23}, requiring a substrate with fixed motor proteins \cite{Tejedor_Ramirez_MAMOL19}, or composing the polymer of active Brownian or Ornstein-Uhlenbeck particles that have the persistence built in \cite{Gompper_Winkler_Active_polymer_analytic_Polymers16,Natali_AOUPhead_polymer_SM20}. 
Here we show, on the contrary, that the active polymer persistence on the chain scale can arise \emph{spontaneously} from isotropic, nonequilibrium, thermal-like fluctuations due to the presence of the topological constraints. Using scaling arguments, supported by simulations, we uncover that the topological restriction of the transverse relaxation modes results in an imbalance of the internal chain tension of entropic origin, which generates the directional dynamics. 

The activity mechanism we use has no persistence and is isotropic. We model it, on each monomer of a contiguous segment of the chain, as thermal-like fluctuations (delta-correlated white noise force with no correlation between the monomers) but with larger amplitude (temperature) than the rest of the chain. Therefore, the active non-equilibrium character of the chain stems from the fact that the two distict segments are maintained at two distinct temperatures. This ``scalar'' activity model is not only interesting from the perspective of materials science, where the stronger fluctuations can be induced by monomer-specific external fields, but also for its relation to chromatin dynamics. The active processes of chromatin (remodelling, transcription, loop extrusion, transcription factor binding, etc) are mediated by the corresponding molecular machinery and exert non-equilibrium fluctuating forces on specific sections of the chromatin fiber \cite{Misteli_genome_structure_function_relation}. The activity, some akin stronger-than-thermal fluctuations \cite{Zidovska_PNAS13,Bruinsma_BPJ14}, impacts the chromatin dynamics \cite{Zidovska_PNAS13,Zidovska_NatComm24} and its conformation \cite{Ashwin_PNAS19,Kardar_active_chromatin_PNAS23,ChanRubinstein_PNAS24,ChanRubinstein_PNAS23}, thereby acting on the biological functionality of the cell \cite{Misteli_genome_structure_function_relation}. Although the role of active fluctuations in chromatin folding is not yet settled, the topological constraints have been shown to govern its large-scale organization \cite{Rings_and_chromosomes_review}. Our present work indicates that the explicit consideration of both the active processes and the chain non-crossability, can be crucial in interpreting the chromatin dynamics, as in synergy with the activity the constraints rectify the chains' motion.

In free space, the center of mass of such a polymer with an active (hot) segment exhibits equilibrium-like diffusive dynamics, with an effective diffusion coefficient $D_{\rm eff}\sim k_{B}T_{\rm eff}/\gamma$, where the effective temperature $T_{\rm eff}=(T_{h}N_{h}+T_{c}N_{c})/(N_{c}+N_{h})$ is the average temperature weighted by the number of monomers $N_{h/c}$ in the hot/cold segments. Indeed, summing up all the Langevin equations for all the $N$ monomers 
\begin{eqnarray}
\label{eq:lan}
m\ddot{\bm{r}}_{i} = -\gamma \dot{\bm{r}}_{i} -\nabla_{\bm{r}_i}U + \sqrt{6 k_{B}\gamma T_{i}} \bm{\eta}_{i},
\end{eqnarray} 
with $\bm{\eta}_{i}$ being uncorrelated random forces of zero mean and unit variance, we get a Langevin equation for the center of mass $M\ddot{\bm{r}}_{\rm cm} = -\Gamma \dot{\bm{r}}_{\rm cm} + \sqrt{6 k_{B}\Gamma T_{\rm eff}} \bm{\eta}$. The potential forces from the monomers ($-\nabla_{\bm{r}_{i}}U$) add up to zero, $M=Nm$ is the total mass and $\Gamma = N\gamma$ the total friction coefficient. A more general model, the so-called Brownian inchworm \cite{Baule_2008}, was solved in an overdamped limit for a dumbell, showing that the drift of the center of mass can arise only as a result of a stretching-dependent friction, which is not present in our model. 

The above equilibrium-like diffusive dynamics in free space is in sharp contrast with that of the linear chain or a ring with active segment in a melt \cite{ATG_NatComm20,ATG_PRR20}. There, the hot segment pulls the rest of the chain through the mesh of the other chains, forcing it to move along the delineated curvilinear path. 
The scaling of mean squared displacement $g_{1}(t)$ of a tagged passive monomer as a function of time is then determined by the chain statistics, namely the scaling of the path's end-to-end distance $R$ with its contour coordinate $s$ as $R\sim s^{\nu}$. For ballistic motion ($s\sim t$) along the path, $g_{1}(t) \sim R^{2}(s(t)) \sim t^{2\nu}$ results, as we detail later below. Therefore, the monomer dynamics is superdiffusive for $\nu>1/2$, and the same was also found for the center of mass mean squared displacement $g_{3}(t)$ of the chains in active topological glass \cite{ATG_NatComm20}. In contrast, in equilibrium the chain also moves within the effective tube formed by the other uncrossable chains, but only diffusively, which results in the well-known subdiffusive reptation dynamics \cite{Edwards_first_Tube,Doi_Edwards_book}. 

Below we first show the topological origin of the superdiffusive dynamics. Then we discuss the dynamical consequences of the directionality on the motion of a monomer and the chain's center of mass. We show how the scaling exponents of the mean squared displacement with time depend on the chain parameters and the spacing of the topological constraints. We conclude with the discussion of the relation to other observed effects in two-temperature polymer melts, chromatin and active-copolymer viscoelasticity.

\section*{Results}
\subsection*{Topological entropy imbalance generates polymer drift}
\begin{figure}[htb]
\centering
\includegraphics[height=5cm]{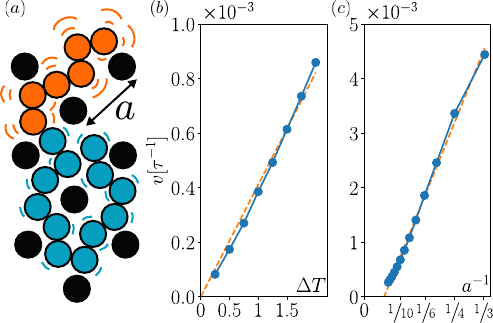}
\caption{\small $(a)$ A 2D sketch of the simulated polymer with a hot segment at temperature $T_{h}$ (orange) and a cold segment at a temperature $T_{c}$ (blue) in a lattice of impenetrable static obstacles (black) forming a square mesh with openings of size $a$ and coordination number $z=4$. $(b)$ and $(c)$ Contour velocity in units of lattice constant per $\tau$ as a function of the temperature contrast $\Delta T$ $(b)$ for $a=9$, and as a function of the inverse lattice spacing $1/a$ $(c)$ for $T_{h}=3$. Dashed lines are linear fits to the data (symbols).}
\label{fig:contour_velocity}
\centering
\end{figure}

A 1D confinement of a two-temperature chain in a tubular channel does not generate the superdiffusive motion observed in \cite{ATG_NatComm20}, because the center of mass dynamics is governed by the same Langevin equation \eqref{eq:lan}. Yet, even in equilibrium there is a distinction between a chain confined in a channel and one confined by the surrounding chains, and it lies in its primitive path $L$. The primitive path is the end-to-end contour length modulo the transverse excursions that are not topologically wrapped around 
the other chains i.e. could be reeled in from the chain ends. The free energy (of entropic origin) cost to stretch a Gaussian chain to end-to-end distance $R$ scales as $F_{s} \sim k_{B}T R^{2}/2N b^{2}$, or equivalently in terms of the primitive path $L$ as $F_{s} \sim k_{B}T L^{2}/2N b^{2}$, $b$ being the monomer size \cite{Rubinstein_book}. In free space or confinement the free energy is minimal at zero stretch, while in practice the fluctuations of order $k_{B}T$ give the random walk statistics $R\sim L \sim b N^{1/2}$. As is well known, in the presence of other chains the situation is different \cite{Rubinstein_book}. While the end-to-end distance remains mostly unaffected (Flory screening of the excluded volume), the primitive path length is strongly stretched. A mean-field picture of the melt is that of a chain in a lattice of impenetrable obstacles with some spacing $a$ and coordination number $z$ (Fig.~\ref{fig:contour_velocity}a). The effective lattice constant $a$ represents the so-called entanglement tube diameter corresponding to the spatial distance between two entanglements that are restricting the chains' motion. The scale $a$ is sensitive to the density and the bending stiffness of the polymer system \cite{Uchida_JCP08,Everaers_MAMOL20}, and it can be much larger than the monomer size, or inter-chain distance because not every contact restricts the chain significantly. At every opening (``gate'') between the obstacles (entanglements) the chain end has $z-1$ options to stretch its primitive path, but only one to return and shorten it. Therefore, this ``topological'' entropy, originating from the broken translational invariance, favors larger $L$, approximately $F_{t} \sim - k_{B}T L/a$, where we dropped factors of order unity involving $z$. Minimizing the total $F_{s}+F_{t}$ gives the standard result of the highly stretched primitive path length, $L\sim b^2 N/a$ \cite{Rubinstein_book}. The stretching is the consequence of the entropic force $f = - \text{d} F_{t}/\text{d} L \sim  k_{B}T /a$ pulling the chain at both ends. 
An exact result of the entropy of the primitive path is known \cite{Helfand_Pearson_JCP83} and shown that the above ``harmonic'' approximation is accurate.

For a chain in a lattice of obstacles, the entropic forces at the two ends, with two distinct temperatures, do not compensate each other, resulting into the force difference 
\begin{equation}
\label{eq:df}
\Delta f \sim \frac{k_{B} \Delta T}{a},
\end{equation}
between the chain ends. Although the system is inherently out of equilibrium and the definition of entropy is problematic, the two ends have locally well defined temperatures and can be described by two competing entropies and the resulting forces. The force imbalance \eqref{eq:df} causes a drift velocity $v\sim \Delta f/\Gamma \sim  k_{B}\Delta T/\Gamma a$ of the monomers along the tube delineated by the obstacles and the contour of the chain. 

To test the theory we run molecular dynamics simulations of a di-block fully-flexible linear chain with identical purely repulsive potentials \cite{KG_model}, where the two blocks differ only by the associated Langevin thermostats at different temperatures (see Methods for details of the model). 
We place the chain in a 3D cubic mesh of impenetrable obstacles, with the lattice constant $a$ (see Fig.~\ref{fig:contour_velocity}a for a 2D sketch), in the range $3\leq a \leq 15$ (in units of the polymer model - see Methods), representing other chains in a melt system. We do so to avoid other, higher-order effects such as the temperature-dependence of the local density/friction, the number of monomers between entanglements $N_{e}$ or active-passive (micro)phase separation \cite{Joanny_Grosberg_PRE15,Ganai_Sengupta_Menon_NAR14,Awazu_active_polymer_separation,Frey_PRL,Smrek_Kremer_PRL17}. We aim to measure the dependence of the contour velocity on the temperature contrast $\Delta T$ and the lattice spacing $a$. 

At first, we equilibrate the chain at $k_{B}T_{c}=1.0$ (see Methods) for all the $N=200$ monomers. Subsequently, we switch the activity of the first $N_{h} = 24$ monomers on by coupling it to the Langevin thermostat $T_{h}>T_{c}$ and keep the rest of the chain with thermostat at $T_{c}$. We let the system run and discard the initial period of $10^{5}\tau$ until the gyration radius of the chain reaches a steady state (about $5\times 10^{4}\tau$). We checked that the total run of over $1.5\times10^{6}\tau$ is long enough for the chain to move more than $8$ times its steady-state gyration radius and reach a diffusive regime in all the systems. For each snapshot (separated by $500\tau$) we determine the effective tube of the conformation by assigning a resident cell to each monomer and excluding the cells belonging to transverse fluctuations (i.e. when the chain leaves a cell and returns back without going around any obstacle). Comparing the tubes in two snapshots separated by $\Delta t$ we determine the surviving part of the tube and for each monomer from the surviving part we compute the displacement along the tube. Averaging over snapshots gives the contour displacement profile $d_{c}(s)$ along the chain confirming a steady contour velocity of the middle section of the chain in the direction of the hot head (Fig.\ref{fig:SI_v_vs_Dt}). 
The mean contour velocity $v$ (in the units of number of lattice cells per $\tau$), computed as $v = \langle d_{c}(s)/\Delta t\rangle_{s\in[75,125]}$, is independent of the different time lags used ($\Delta t = 500- 2000\tau$) (see Supporting Information (SI) in Appendix \ref{sec:SI} Fig.~\ref{fig:SI_v_vs_Dt}). Simulations for different $\Delta T=T_{h}-T_{c}$ and different $a$ confirm the linear scalings $v\sim \Delta T$  as well as $v\sim a^{-1}$ (Fig.~\ref{fig:contour_velocity}b,c). While in the former case, the linear fit with a single parameter is equivalent to a two-parameter fit, confirming $v(\Delta T=0)=0$, the $v\sim a^{-1}$ requires a two-parameter fit 
\begin{align}
\label{eq:v_finite}
v \sim \frac{\Delta T}{a} - \frac{\Delta T}{a_{\rm max}},
\end{align}
where we find $a_{\rm max}=18.2$, which is comparable to the chain size characterized by its gyration radius $R_{g} \simeq 12.5\sigma$ or the end-to-end distance $R_{ee}\simeq 29.5\sigma$ (both mostly independent on $a$ - see Fig.~\ref{fig:SI_Ree_Rg}).
Clearly, the theoretical result \eqref{eq:df} holds when the chain is much larger than the lattice spacing $a$, while if the chain is smaller than the lattice spacing, there are no topological constraints and we expect $v \to 0$.

\begin{figure}[htb]
\includegraphics[height=5.5cm]{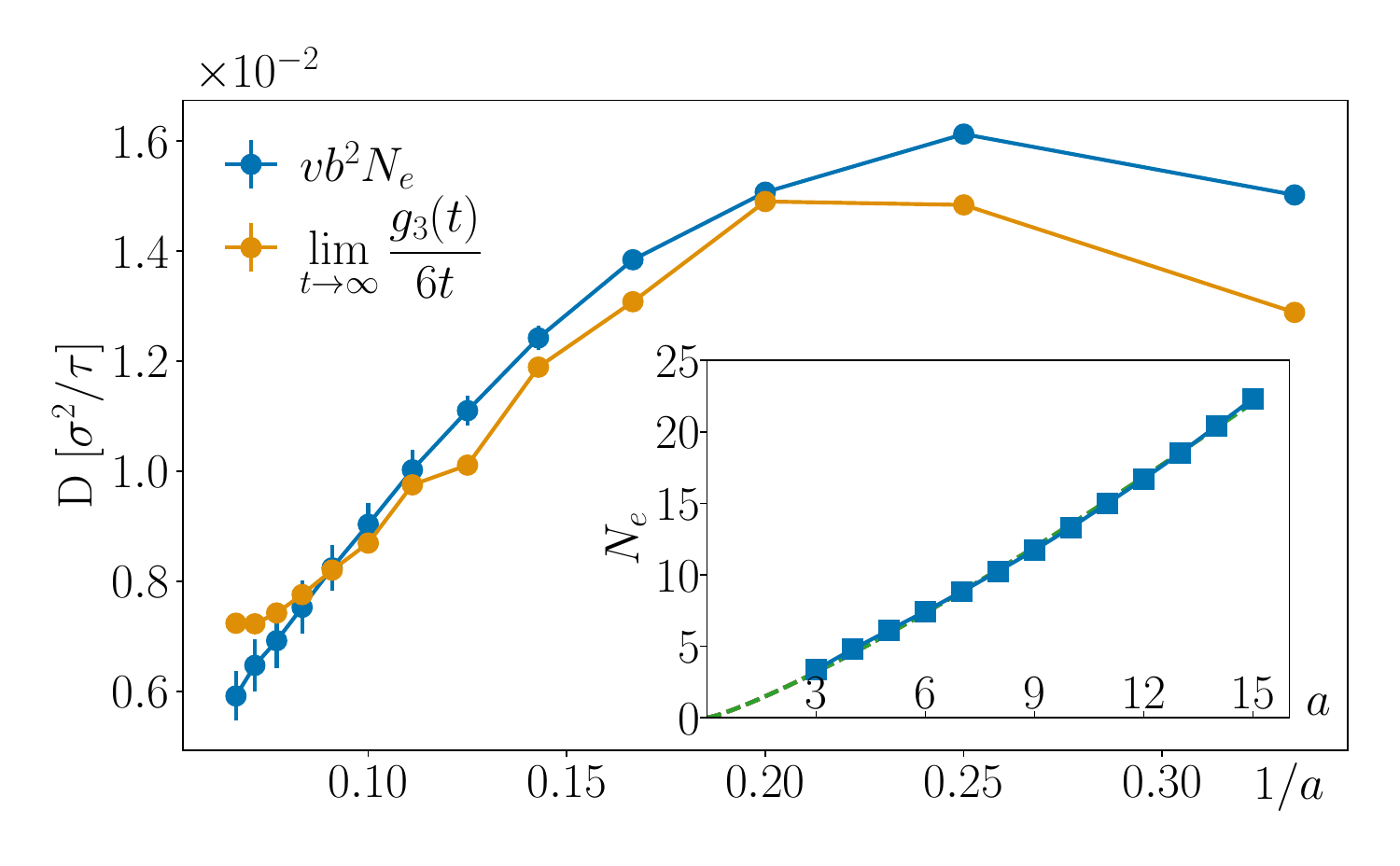}
\vspace{-1cm}
\caption{\small  Diffusion coefficient as a function of the inverse lattice spacing $1/a$ for $T_{h}=3.0$ measured from the contour velocity as $v b^{2} N_{e}$ (blue) or directly from the long time mean-squared displacement (orange). Inset: Number of monomers per one cell of the primitive tube, $N_{e}$, as a function of the spacing $a$ (symbols). The fitted power-law (green dashed) has the form $N_{e}=0.85 a^{1.19}$.}
\label{fig:Deff}
\centering
\end{figure}

We can verify these results also by computing the diffusion coefficient of the whole chain. The time to reach a completely uncorrelated conformation, the relaxation time $\tau_{\rm relax}$, is the time it takes for the chain to get out of its initial tube \cite{Edwards_first_Tube}. The length of this primitive path/tube can be also written as $L\sim a N/N_{e}$, where $N_{e}$ is the number of monomers per one cell. For long chains, the number of cells forming the primitive path $L/a =  N/N_{e} \gg 1$) the drift will dominate over diffusion of the chain and, therefore, $\tau_{\rm relax} \sim (N/N_{e})/v$.  On that time scale, the chain starts to diffuse and moves its own end-to-end distance $R\sim N b^{2}$ in the 3D space, so the diffusion coefficient $D\sim R^{2}/\tau_{\rm relax} \sim v N_{e} b^{2}$. Comparing the latter relation with $D$ computed from the long-time limit of the mean-square displacement of the center of mass $D = \lim_{t\to\infty}g_{3}(t)/6t$ is a direct way to check the result \eqref{eq:df}. Such a confirmation is indeed independent as the long-time limit dynamics is juxtaposed with the contour velocity $v$ that is extracted from short time scales as explained above. In Fig.~\ref{fig:Deff} we plot together $v b^{2} N_{e}$ and $D$ as a function of $1/a$ and observe an agreement without any fitting parameters. The used dependency $N_{e}(a)$ (Fig.~\ref{fig:Deff} inset) we measured as $N/(L/a)$.
For equilibrium polymers with no excluded volume the polymer follows random-walk statistics $a\sim b N_{e}^{1/2}$, and in the case with excluded volume $a\sim b N_{e}^{\nu}$, where $\nu$ crosses over from $1$ at small scales (rigid rod) to $0.588$ (self-avoiding walk) at large. We find our data to fit well with $\nu\simeq 0.84$ (Fig.~\ref{fig:Deff} inset). The scaling estimate $D\sim  v N_{e} b^{2}$ captures well the non-monotonic behavior of $D$ at low $a$ (increase of $D$ from $a=3$ to $a=4$) and then decrease for higher $a$. Both facts result from the fine interplay of the finite size changes of $v$ and $N_{e}$ as a function of $a$. The regime of high $a\geq 13$ is not captured by the scaling estimate accurately suggesting that for slow drifts (large $a$), diffusion along the primitive tube might be at play.
Beyond the scaling arguments, the classical tube reptation theory \cite{Doi_Edwards_book} can be extended by a drift term and is then governed by a drift-diffusion equation \cite{Tejedor_Ramirez_MAMOL19,Doi_Edwards_book}, capturing the crossover from drift-dominated to diffusion-dominated diffusion coefficient. As we show in the SI (sec.~\ref{sec:SI_coth}) all our systems are purely drift-dominated and the leveling off of the $D$ at large $a$ is rather caused by the transverse relaxation.

The entropic drift generates a distinctive dynamics of segments or their center of mass, relevant and accessible also experimentally in the context of chromatin fiber in living cells. Given the isotropy of the active entropic pulling, the center of mass performs diffusion on long time scales ($t>\tau_{\rm relax}$). However, below the relaxation time, the center of mass and the monomer dynamics can exhibit both subdiffusive and superdiffusive behaviors, depending on the activity of the segment and the examined length and time scales. In rough terms, on small length scales and short times the dynamics is Rouse-like (subdiffusive for a monomer), although modulated by the higher effective temperature. On longer scales, the entropic pulling starts to play a role and a superdiffusive regime emerges. On yet longer scales a monomer performs ballistic motion over a curvilinear trajectory coupling the static conformational exponent with the dynamic one characterizing the mean-squared displacement before the diffusive regime settles at long times.

\subsection*{Segmental superdiffusive dynamics is coupled to the chain conformation}
\begin{figure*}[htb]
\includegraphics[height=5.5cm]{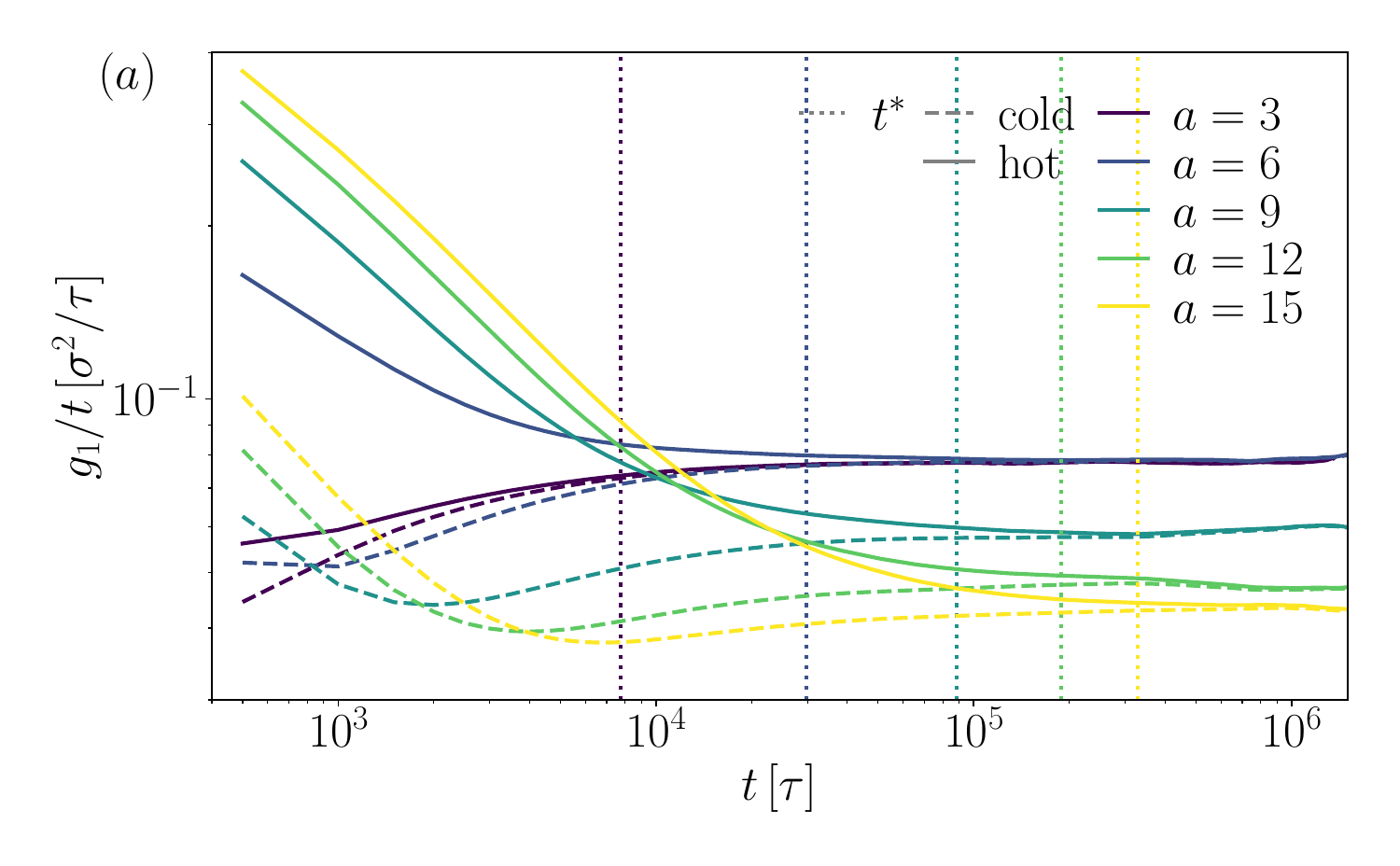}
\includegraphics[height=5.5cm]{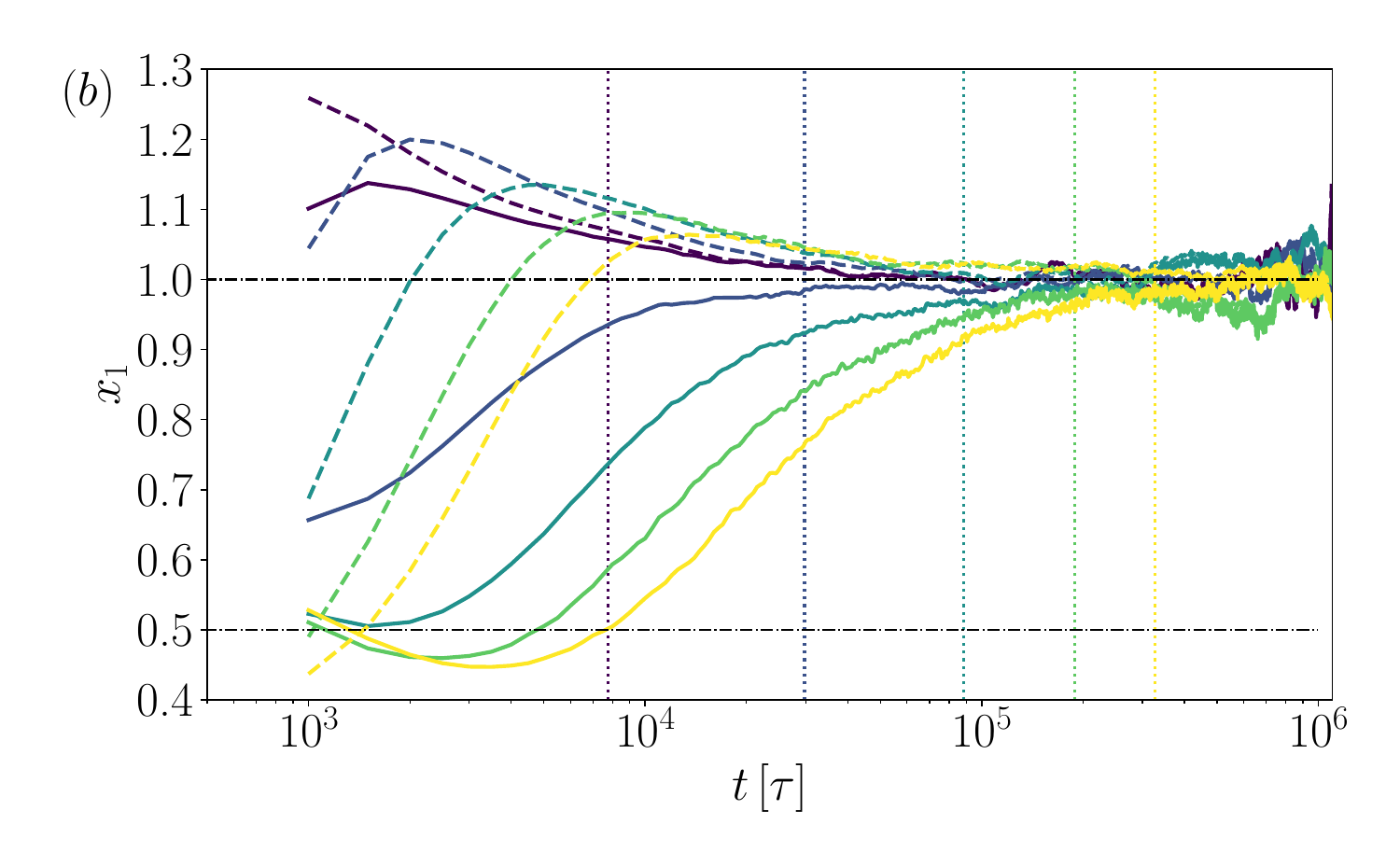}\vspace{-0.5cm}
\includegraphics[height=5.5cm]{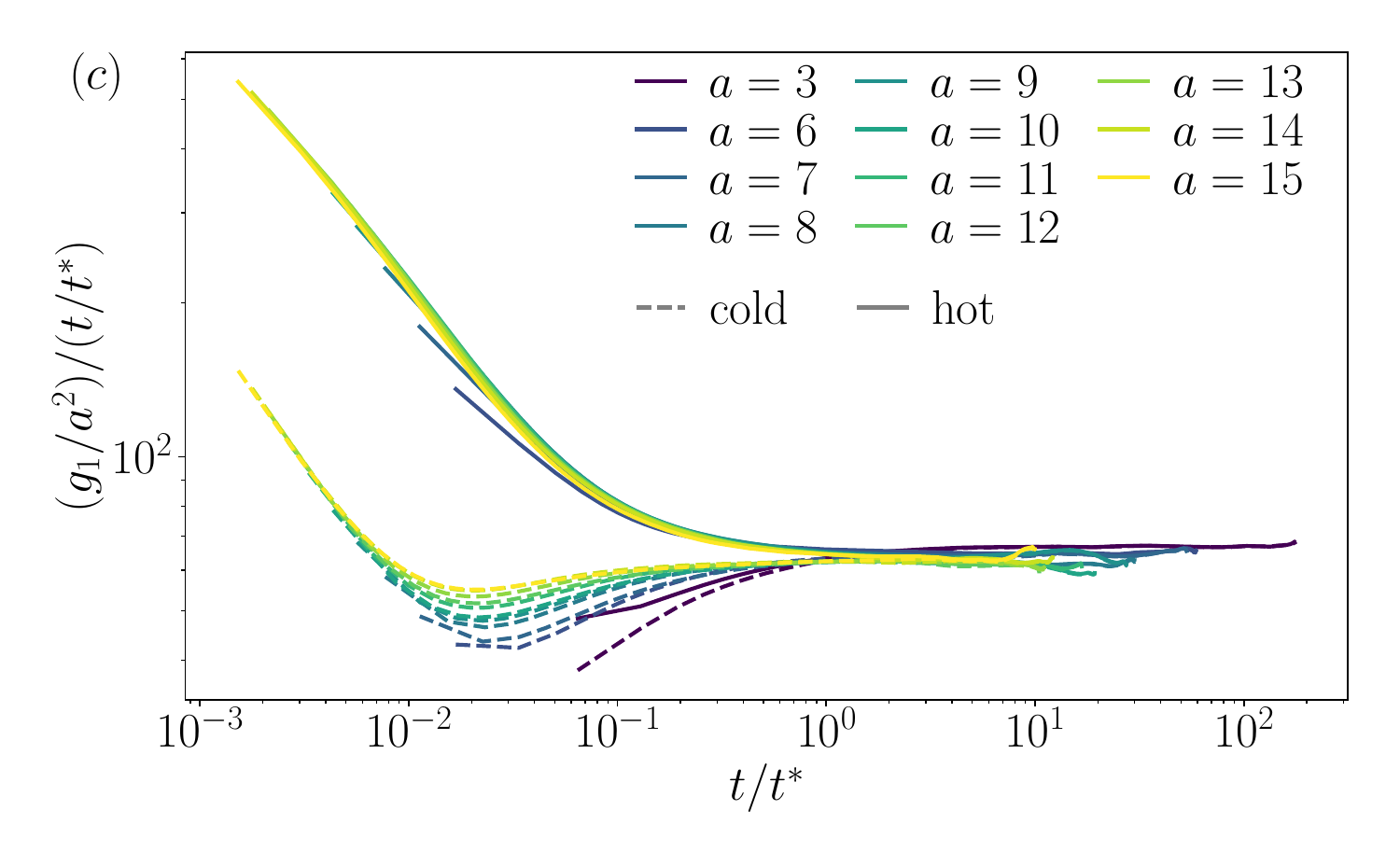}
\includegraphics[height=5.5cm]{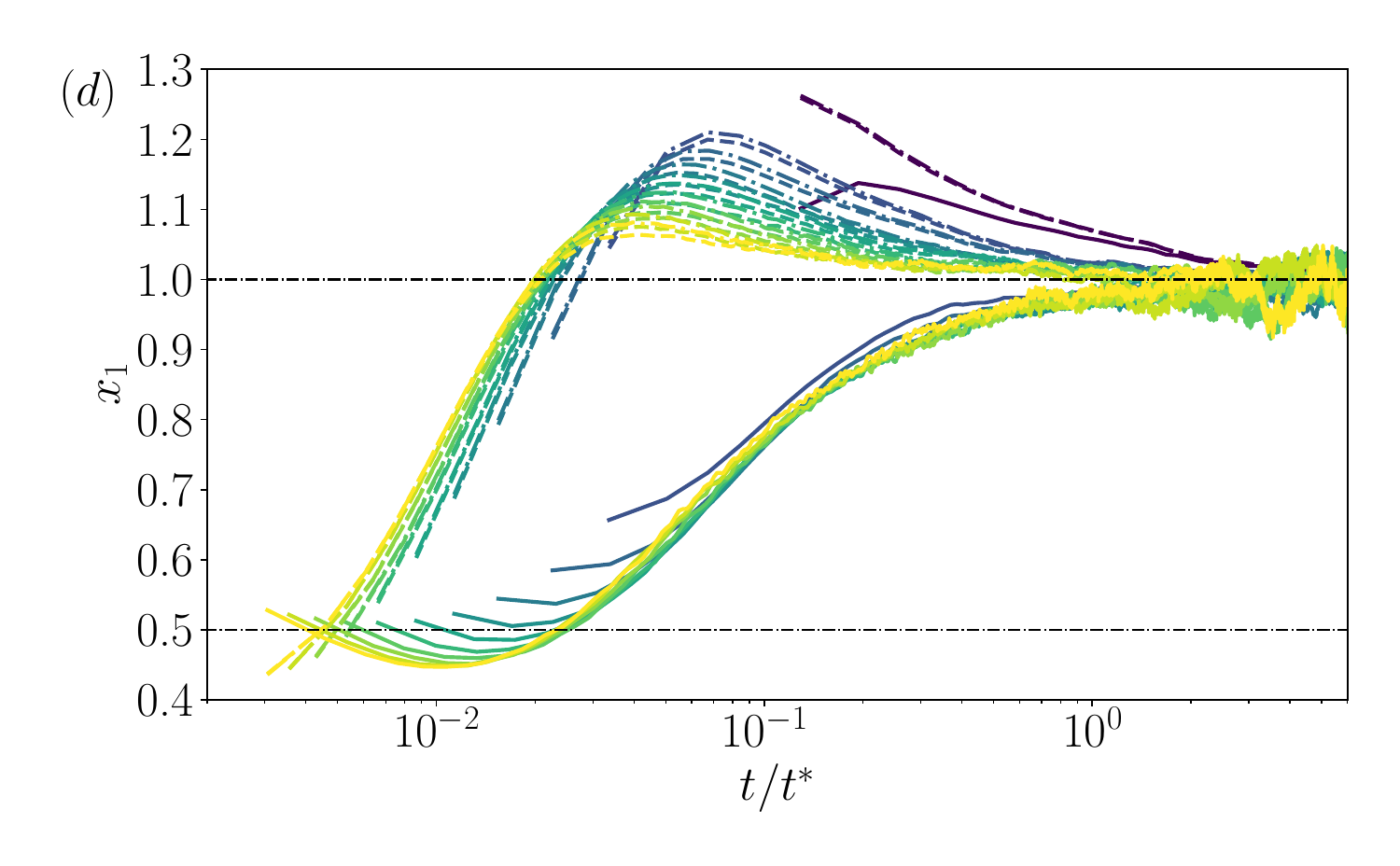}
\caption{\small $(a)$ Mean square displacement divided by time $g_{1}(t)/t$ of a monomer in the middle of a hot/cold (solid/dashed) segment as a function of time for different lattice spacing $a$ (colors). $(b)$ Effective scaling exponent $x_{1} = \text{d}\ln(g_{1})/\text{d}\ln(t)$ of the mean squared displacement. Dot-dashed lines highlight exponents $1/2$ (Rouse) and $1$ (diffusion). $(c)$,$(d)$ The same data as in $(a)$ and $(b)$ respectively normalized by $a^{2}$ and $t^{\ast}$ (see text for definition, dotted) as marked in the axes labels. For clarity, fewer systems are shown in the non-normalized [$(a)$,$(b)$] version.}
\label{fig:msd}
\centering
\end{figure*}

To see the general trends discussed above in more detail, we measure the mean-square displacements $g_{1}(t)$ of the central monomer of the hot and cold segments respectively as in \eqref{eq:g3_time_avg}. In the long-time limit $g_{1}/t$ plateaus for all $a$ and both segment types, signifying ordinary diffusion (Fig.~\ref{fig:msd}a). At short times the hot segments in systems with large $a$ are subdiffusive, consistent with Rouse $g_{1}\sim t^{x_{1}}$ with $x_{1}=0.5$ (Fig.~\ref{fig:msd}a,b). 
The active dynamics sets on when the hot segment explores several cells and, as a result, the cold segment is pulled and exhibits superdiffusive dynamics ($x_{1}>1$ in $g_{1}(t)\sim t^{x_{1}}$). In contrast to systems with large $a$, the system with $a=3$ is superdiffusive already at the smallest time resolution $500\tau$. This is true not only for the cold segment as expected, but also for the hot segment. The latter is caused by the fact that it is the chain tip that creates the entropic pulling force, while for such a small mesh size, the hot segment spans several cells and its midpoint already experiences the pulling of the tip. If we track the $g_{1}$ of the hot tip, we recover its subdiffusion (see SI Fig.~\ref{fig:SI_g1_hot_head}). The case of $a=3$ shows that an \emph{internal} hot monomer can exhibit transient superdiffusion, which confirms that the pulling is generated at the tip and corroborates the entropic origin of the directionality. 

We can recover a universal segmental chain dynamics if we coarse-grain the dynamics to the lattice scale, by normalizing the $g_{1}$ by $a^{2}$ and the time scale by a proper $t^{\ast}(a)$. The time scale $t^{\ast}(a)$ has to be larger than the relaxation time and such that in all systems the cold segment explores the same number of cells. We find a good collapse (Fig.~\ref{fig:msd}c) of the $g_{1}$ curves if $g_{1}(t^{\ast}(a))$ of the cold head of the chain reaches at least $(8a)^{2}$, which agrees with the number of cells ($\simeq 60$, SI Fig.~\ref{fig:SI_tube_length}) corresponding to the chain in the system with $a=3$. 

The short-time superdiffusive regime $x_{1}(t)>1$ (Fig.~\ref{fig:msd}b,d) stems from the fact the monomer moves ballistically along the delineated (primitive) path that exhibits non-ideal statistics ($2\nu>1$) at these scales. In the artificial case of infinitely long monofractal chain, characterized by $2\nu$ independent of the scale $s$, the exponent  $x_{1}=2\nu$ would be a constant independent of the time scale. In reality, however, the superdiffusive exponent $x_{1}$ and its time-dependence are affected by three finite-scale effects: $(i)$ the mesh length scale $a$, defining the onset of entropic pulling, $(ii)$ total chain length above which the system loses correlation and hence becomes diffusive and $(iii)$ the scale dependence of the effective exponent $2\nu(s)$ that characterizes conformational features of the chain across the scales. 

\begin{figure}[htb]
\includegraphics[height=5.5cm]{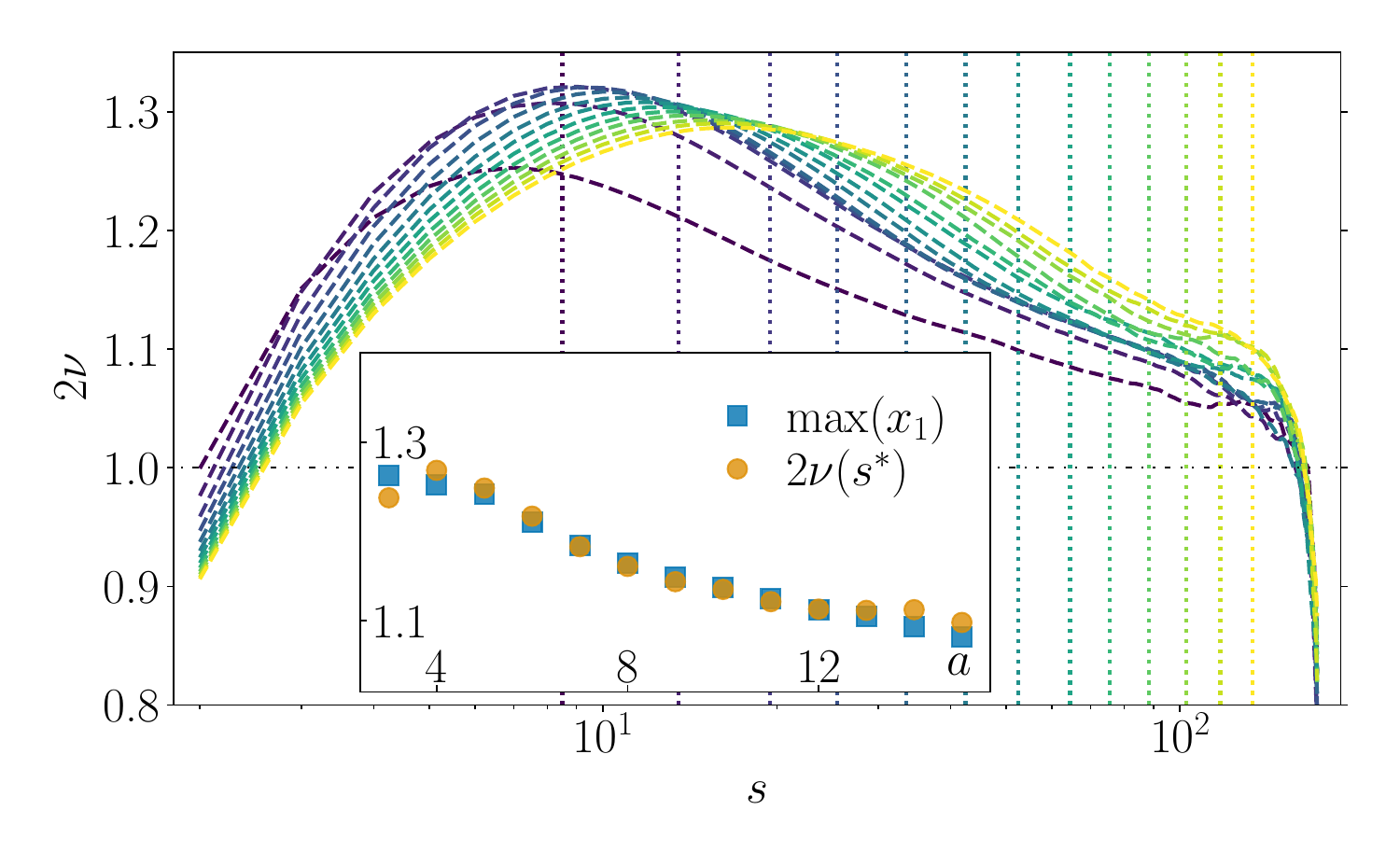}
\caption{\small The exponent $2\nu$ of the cold segment as a function of the contour distance $s$. The vertical dotted lines are at $s^{\ast}$ (see text). Inset: Comparison of the exponents $\max(x_{1})$ and $2\nu(s^{\ast})$. Note that the value of $\max(x_{1})$ for $a=3$ is underestimated because of the low sampling frequency.}
\label{fig:x1_vs_a}
\centering
\end{figure}

To determine their effects, we extract $2\nu$ from the scaling of the mean squared internal distance $d^{2}(s) = \langle(\bm{r}_{i}-\bm{r}_{i+s})^{2}\rangle$, where the average is over the monomer contour index $i$, and from there the effective exponent $2\nu(s) = \text{d}\ln(d^{2}(s))/\text{d} \ln s$ (Fig.~\ref{fig:x1_vs_a}). The $x_{1}$ can reach the value of the maximum of $2\nu(s)$, which is almost the case for low $a$, but for higher $a$ it reaches only lower values because the effects of the transverse fluctuations and the finite chain length are more pronounced. As the superdiffusion emerges only on scales above $a$, the exponent $x_{1}$ achieves its maximum only when the monomer moves at least $k\geq 1$ cells. At that scale ($k a$) the effective exponent $2\nu$ is lower and correspondingly lower is also $x_{1}$. To see that, we find $s^{\ast}$ where $d^{2}(s^{\ast})=(k a)^{2}$, compare $2\nu(s^{\ast})$ with $\text{max}(x_{1})$ and find a good agreement when $k^{2}=3$ (Fig.~\ref{fig:x1_vs_a} inset). For a general driving velocity $v$, the scale $k$ would depend on $v$ and would be given as the scale along the tube above which the drift dominates the diffusion. In our case of entropic driving the drift is coupled to the contour displacement and therefore the criterion of constant $k$ is applicable. More important than the actual fitting value of $k$ is that other choices of $k$ reproduce the dependence of ${\rm max}(x_{1})(a)$ up to a constant shift (SI Fig.~\ref{fig:SI_x1_max_vs_a_vs_k}), indicating the presented connection of $2\nu$ and $x_{1}$ and that it is the finite length and time variation respectively generating the diversity of the exponents. In contrast to \cite{Tejedor_Ramirez_SM20} where the disappearance of a superdiffusive regime for large number of entanglement units $Z=N/N_{e}\sim a^{-2}$ has been seen, we observe here the opposite trend. For smaller $a$ the superdiffusion is more pronounced, precisely because $2\nu>1/2$ is more apparent when a finite chain occupies more ``entanglement'' cells.

\subsection*{Ballistic dynamics of the center of mass is limited by the chain length}

\begin{figure*}[htb]
\includegraphics[height=5.5cm]{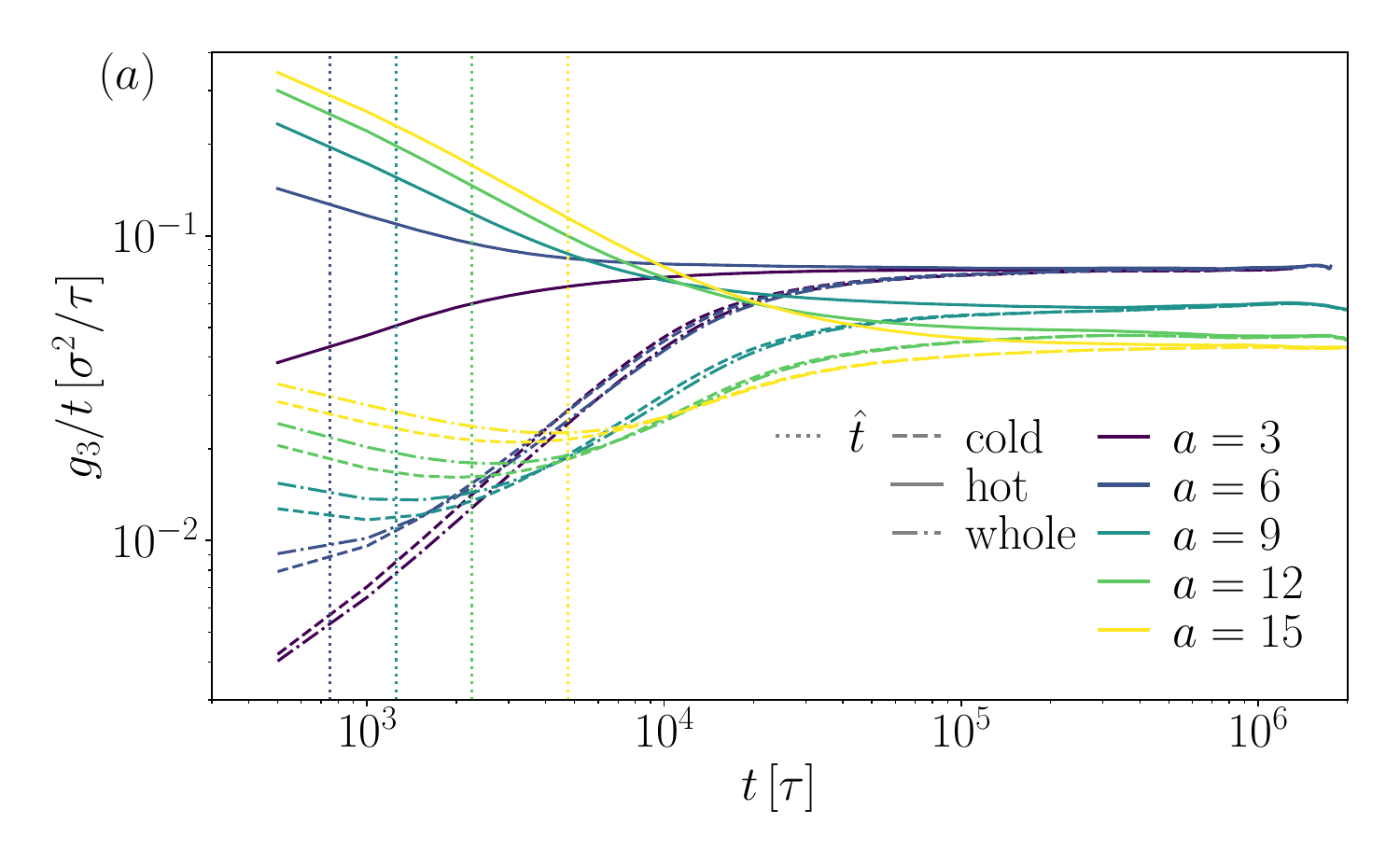}
\includegraphics[height=5.5cm]{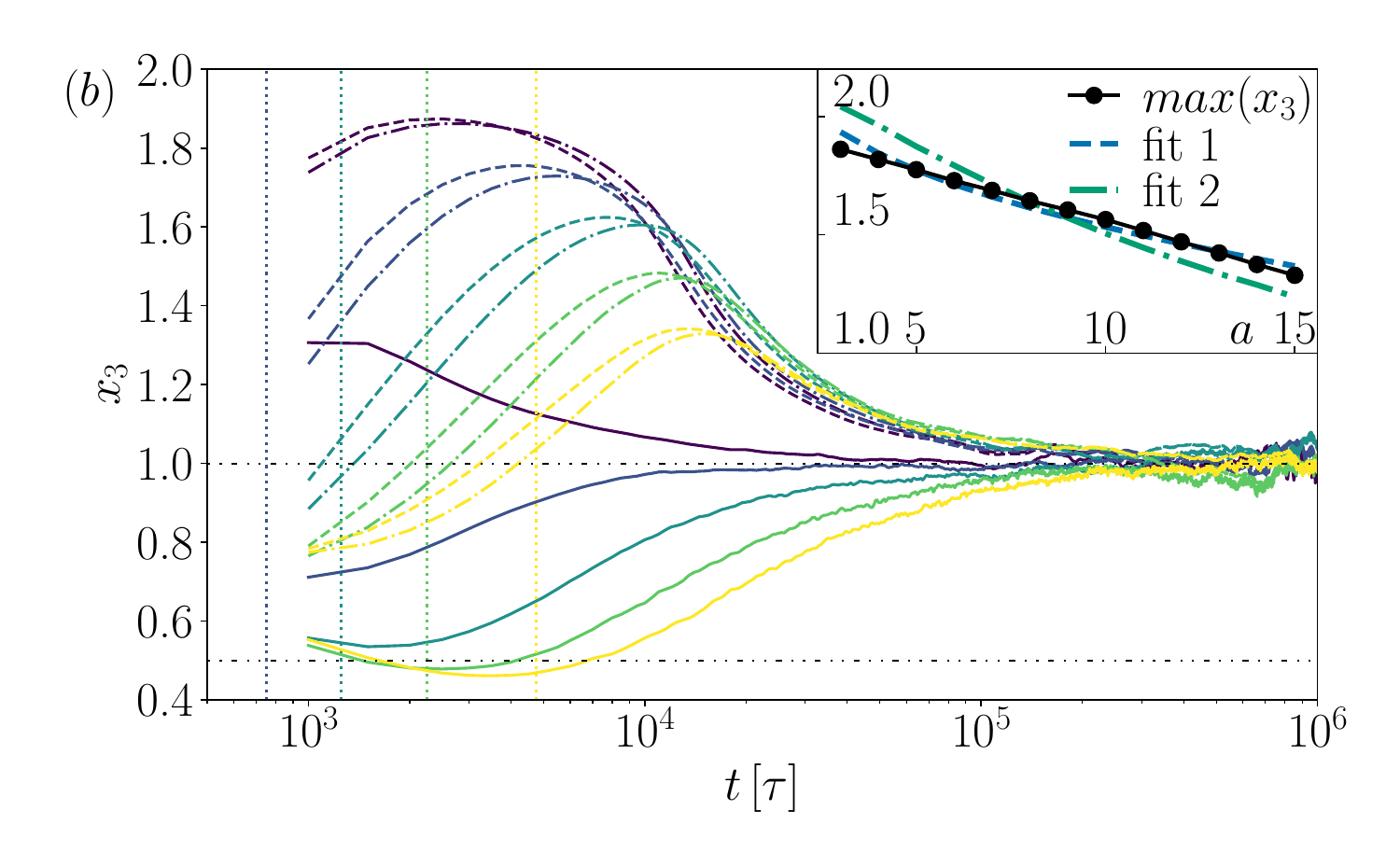}\vspace{-0.5cm}
\includegraphics[height=5.5cm]{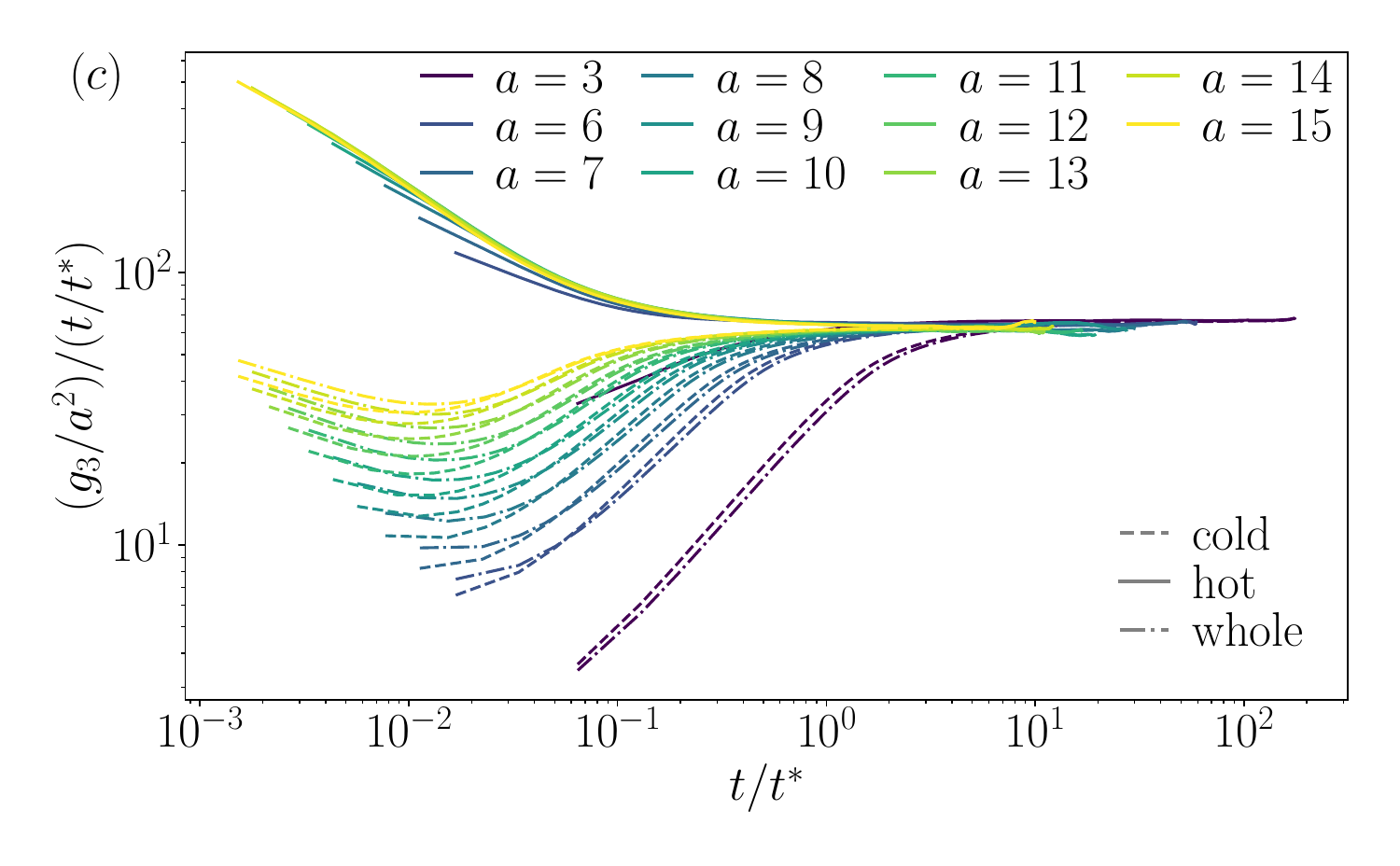}
\includegraphics[height=5.5cm]{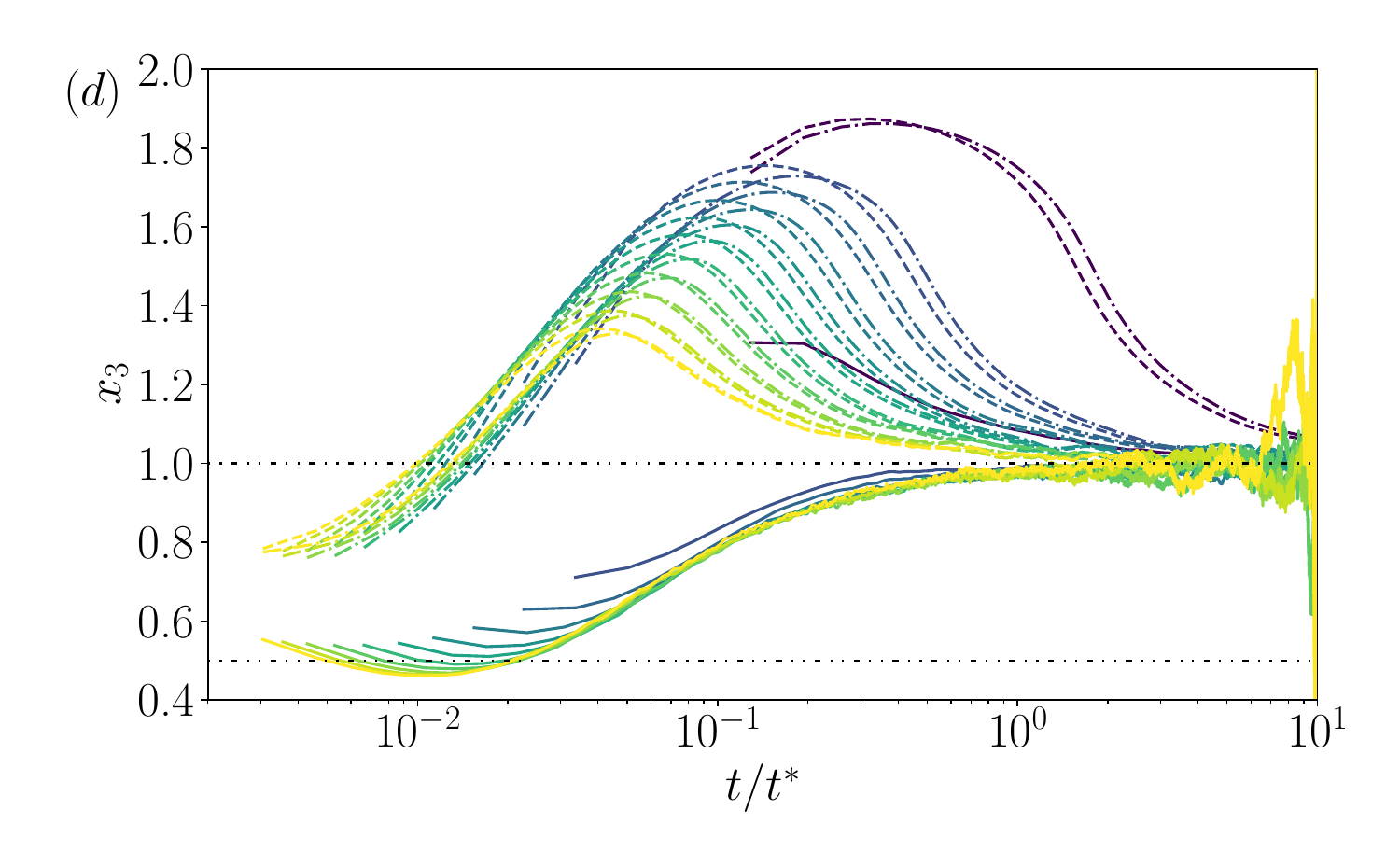}
\caption{\small $(a)$ Mean square displacement divided by time $g_{3}(t)/t$ of a center of mass of the hot/cold/whole (solid/dashed/dot-dashed) segment/chain as a function of time for different lattice spacing $a$ (colors). The vertical lines represent the time $\hat{t}$ when the hot segment's $g_{1}(\hat{t}) = 2.5a^{2}$ signifying the onset of the superdiffusion. $(b)$ Effective scaling exponent $x_{3} = \text{d}\ln(g_{3})/\text{d}\ln(t)$ of the mean squared displacement. Dot-dashed lines highlight exponents $1/2$ (Rouse dynamics for a segment, reptation for the total center of mass) and $1$ (diffusion). Inset: The maximal exponent $x_{3}$ as a function of the lattice spacing $a$ (symbols). Fit 1 is using \eqref{eq:v_finite} while fit 2 directly the measured values of $v(a)$ (see text). $(c)$ and $(d)$ data normalized by $a^{2}$ and $t^{\ast}$ (see text for definition) as in the axes labels. For clarity, fewer systems are shown in the non-normalized [$(a)$,$(b)$] version.}
\label{fig:g3}
\centering
\end{figure*}

The center of mass mean-squared displacement $g_{3}(t)$ of the whole chain, \eqref{eq:g3_time_avg}, exhibits diffusion on long time scales ($t>\tau_{\rm relax}$) as the primitive path of the chain loses orientational correlation. Note the diffusion sets on at about the same absolute time for all $a$ (Fig.~\ref{fig:g3}$(a)$), because the relaxation time $\tau_{\rm relax} \sim (L/a)/v $ and both, the primitive path $L/a$ (SI Fig.~\ref{fig:SI_tube_length}) and $v$ are approximately proportional to $a^{-1}$.  At early times, $g_{3}(t)$ shows the transition from sub-diffusive (reptation) 
 to superdiffusive, which sets in at time $\hat{t}$ (Fig.~\ref{fig:g3}a) when the hot segment explores a few cells ($g_{1}(\hat{t})= 2.5a^{2}$). The highest exponent $x_{3}= \text{d}\ln(g_{3})/\text{d}\ln(t)$ is reached when the entropic pulling fully develops ($\simeq 4\hat{t}$). In contrast to tangentially driven filaments where the center of mass typically exhibits ballistic motion $x_{3}=2.0$ for $t<\tau_{\rm relax}$ \cite{Xiang_tangentially_driven_filaments_PRR23,Ubertini_ACSML24} we observe $a$-dependent exponents that arise from the finite chain size and transverse Rouse relaxation on scales smaller than $a$. Indeed, for small $a$, although each cell is explored by monomers quickly, the transverse fluctuations are strongly suppressed by the mesh, thereby the pulling resembles tangential driving. The primitive path of the chain spans tens of cells (SI Fig.~\ref{fig:SI_tube_length} and section \ref{sec:SI_g3}), which together make long enough trajectory for the ballistic regime to develop. In contrast, in the largest cells the segment fluctuations are larger and the whole chain spans only several cells. To see the effect of the number of cells spanning the chain is to compare the time when $g_{3}\simeq a^{2}$ and the time when $g_{3}\simeq R_{ee}^{2}$. The corresponding time span, larger for smaller $a$, coincides with that when $x_{3}$ culminates (SI Fig.~\ref{fig:SI_g3_normalized_by_Ree}). 
 
The fact that the ballistic regime does not have enough time and space to develop before the diffusion sets in is illustrated by the higher $\hat{t}$ for higher $a$'s. Another way to illustrate the idea is when we recover the universal dynamics of the center of mass by coarse-graining to the level of the mesh (Fig.~\ref{fig:g3}c,d, like for $g_{1}$ by normalizing the time by $t^{\ast}$ and $g_{3}$ by $a^{2}$). The initial rise of the exponent $x_{3}>1$ exhibits overlap for systems with different $a$ and lasts longer (and therefore reaches higher values) for systems of low $a$ spanning more cells. More details on the crossover to the ballistic regime and its derivation are in the SI. 

Tejedor and Ram\'{i}rez \cite{Tejedor_Ramirez_SM20}, considering a tangentially driven chain with constant drift on the tube level, explain the superdiffusion exponent $x_{3}$ as a crossover between the short time diffusion dominating over drift and long time diffusion above the chain scale that is governed by the drift. Based on the crossover times between the regimes and the value of $g_{3}$ they derive an approximate expression for $x_{3}\simeq 0.657\log(cZ^{2})+1.77$, where $c$ is the drift in the lattice units per entanglement time $\tau_{e}=\tau_{0} N_{e}^{2}$, with $\tau_{0}$ being the monomer diffusion time, and $Z=N/N_{e}$ is the number of entanglements. In our units, this translates to $x_{3}\simeq 0.657\log(A v)+1.77$, where $A=N \tau/\tau_{0}$ and the only dependence on the entanglement (lattice) scale comes from the velocity $v(a)$. We use this approximation to fit the data of the  ${\rm max}(x_{3})$ reached in the simulation for the given $a$, in two ways. In the first one, we used the expression \eqref{eq:v_finite} for $v(a)$, using $A$ and $a_{\rm max}$ as the fitting parameters, and in the second one we use directly the measured values of $v$ and using only $A$ as a fitting parameter. Both fits capture the main trend (inset of Fig.~\ref{fig:g3}b), but fail to reproduce the measured dependency details as highlighted by the disagreement of the fitted parameter $a_{\rm max}\simeq 42$ with the one obtained from \eqref{eq:v_finite} $a_{\rm max} \simeq 18$. We attribute these differences to explicit transverse fluctuations of the chain that speedup the chain relaxation (SI sec.~\ref{sec:SI_g3}) and were not considered in the analytical model \cite{Tejedor_Ramirez_SM20}, but further work is needed to confirm this hypothesis.

\section*{Discussion}
Besides the present directional dynamics, two-temperature polymer melts can exhibit also the emergence of the active topological glass (ATG) \cite{ATG_NatComm20,ATG_PRR20} and the active-passive (micro)phase separation \cite{Joanny_Grosberg_PRE15,Smrek_Kremer_PRL17}. While the precise interconnections of the three phenomena are not resolved, here we explain the physical origin of the directionality that is a neccessary condition for the occurence of ATG. However, in contrast to the ATG and active-passive phase separation, both of which occur only above a critical temperature contrast $\Delta T>\Delta T^{\ast}$, the pulling mechanism at hand occurs at any positive temperature contrast $\Delta T>0$ as confirmed by $\lim_{v \to 0} \Delta T = 0$ in Fig.~\ref{fig:contour_velocity}. Similarly, while the former phenomena are conditioned on a critical length of the hot segment, which depends on the temperature contrast, the present rectification mechanism, is at play as long as the chain possesses at least one hot monomer and the chain is larger than the lattice spacing. However, we expect that if the hot segment itself is smaller than the lattice spacing, the driving is caused by an effectively lower temperature contrast $\Delta T_{\rm eff}$, because the segmental  dynamics is governed by the coarse-grained Langevin equation \eqref{eq:lan} on the scale $a$. On the scale of the cell the chain segment has the effective temperature $\Delta T_{\rm eff} = T_{{\rm eff},a} - T_{c}$ where $T_{{\rm eff},a}=(T_{h}N_{h,a}+T_{c}N_{c,a})/(N_{c,a}+N_{h,a})$ is the number average temperature of the segment within the cell of size $a$. Therefore $T_{{\rm eff},a}=T_{h}$ only as long as $N_{c,a}=0$.

The situation is more complex if the hot segment is located somewhere in between the chain extrema, which is a relevant case for chromatin model with heterogeneous activity. It has been shown analytically for a model \emph{without} excluded volume or topological constraints that the activity induces effective looping of the active segment and effective attraction of the flanking passive regions \cite{Kardar_active_chromatin_PNAS23}. Although the full model with the excluded/topological interaction is considered analytically inaccessible \cite{Kardar_active_chromatin_PNAS23}, here we show a way to treat it with an effective theory. The entropic pulling of the intermediate hot segment competes with the stretching of the primitive path at the flanking ends, resulting in directional motion of the active loop. As the conformational data on chromatin cannot distinguish passive (segmental segregation due to distinct interactions) from active (due to different activity levels) mechanisms of chromatin folding, the dynamics and kinetics are the decisive aspects to discriminate the two mechanisms \cite{Kardar_active_chromatin_PNAS23}. Recent experiments \cite{Zidovska_NatComm24} on living chromatin investigate the dynamics of two active/inactive gene loci and their correlation with the compaction and the motion of the background chromatin. The active genes displacements are more correlated with the motion of the less compacted background chromatin and the effective scaling exponents of the $g_{1}$ of the genes depends more on chromatin compaction than the activity. Although the exponents (or their time behavior) cannot be measured accurately on the sample, the work shows the important differences in the active/inactive gene dynamics and their dependence on the local properties. Note also that topological mesh in chromatin is itself spatially heterogeneous as the density and stiffness of the fibers varies and therefore large variability can be expected in chromatin experiments on many genes. Here we demonstrate a mechanism that can affect the chromatin folding and dynamics. Already the present model exhibits a large variability in the effective dynamical scaling exponents $x_{1}$ and $x_{3}$ and we specify how these depend on the segment's identity ($h/c$), the local topological mesh size $a$ and the observation time scale. An interesting future step is to investigate the impact of the hydrodynamic interactions that we neglect due to the typical screening effects in polymer melts in equilibrium, but the local flow field around the pulled chain might lead to local coherent motion \cite{Zidovska_PNAS13}.

The active dynamics impacts the viscosity of the active copolymer melt. We expect the scaling of the viscosity with the polymer length to follow that of the tangentially active polymers $\eta \sim N^{2}$ \cite{Suvendu}, however, for a different reason. While the scaling in tangentially driven polymers at every monomer arises from linear scaling of both, the relaxation time $\tau_{\rm relax} \sim N$  \cite{Tejedor_Ramirez_Pol23}, and the corresponding energy density of the active forces scale as $G_{0}\sim N$ \cite{Suvendu}, giving $\eta\sim G_{0}\tau_{\rm relax}$, in the two-temperature polymers is the $\tau_{\rm relax} \sim \Gamma L/ v \sim N^{2}$, while $G_{0}$ would not scale with length unless the hot segment is proportional to the total length as well. These estimates rely however on the assumption that the entanglement effects are similar to the equilibrium case and independent of the driving, which might not be generally satisfied. Specifically, one assumption in this context we use and was also used in directional reptation theory \cite{Tejedor_Ramirez_MAMOL19} is that the mesh size and the tube length are independent of the driving. The driving on both ends stretches the chain which was shown to impact the properties \cite{Li_SM24}. 

The present work explains how the broken translational symmetry due to (even temporary) topological constraints together with heterogeneous fluctuations generate directional motion of active-passive copolymers. We present how to understand these systems analytically using the language of equilibrium statistical mechanics expanded to several entropies associated to different thermostats. The arising directional motion  will likely affect steady state phases and the kinetics of both active-passive or genuine chemical copolymer melts. Exploring these directions further will bring deeper understanding of living chromatin and promises the development of functional active polymeric materials.

\section*{Acknowledgement}
The computational results presented have been achieved using the Vienna Scientific Cluster (VSC).

\section*{Contributions}
JS conceived the study and formulated the theory. AH, IC and JS ran the simulations and analyzed the data. All the authors interpreted the data, JS wrote the manuscript with contributions from all the authors.

\appendix
\section{Methods}
\subsection*{MD model}
We use the stadard Kremer-Grest model with no bending rigidity. All beads interact with WCA potential
\begin{equation}
U_{\mathrm{WCA}}(r) = 4\varepsilon \left[ \left(\frac{\sigma}{r}\right)^{12} - \left(\frac{\sigma}{r}\right)^{6}\right] +\varepsilon,
\label{eq:LJ}
\end{equation} 
for $r<2^{1/6}\sigma$ and $0$ otherwise, where the scale $\sigma$ represents the bead (monomer) diameter $b$. The mesh of obstacles is formed by immobile beads placed side-by-side (centers separated by $\sigma$) arranged in lines of a cubic grid with $a$ beads per cell side. When referring to the grid cell size $a$ as lattice spacing we use the units of $sigma$, e.g.~$a=3$ means $a=3\sigma$. The monomer beads interact with the mesh beads with the same potential \eqref{eq:LJ}. The polymer connectivity is given by the finitely extensible nonlinear elastic potential
\begin{equation}
U_{\mathrm{FENE}}(r)  = -\frac{1}{2}r_{\mathrm{max}}^{2} K \log\left[1-\left(\frac{r}{r_{\mathrm{max}}}\right)^{2}\right],
\label{eq:FENE}
\end{equation}
where $K=30.0\varepsilon/\sigma^{2}$ and $r_{\rm max}=1.5\sigma$, making the chains essentially uncrossable. The potential parameters, together with the bead mass $m=1$ set the time scale $\tau = (m\sigma^{2} /\varepsilon)^{1/2}$. Given the parameters this is of the same order as the monomer diffusion time $\tau_{0}=b^{2}\frac{\gamma}{k_{B}T}$. All polymer beads are connected to their respective Langevin thermostat at temperature $T_{h/c}$, depending on if they are hot or cold, while both have a coupling/friction parameter $\gamma=2/3\tau^{-1}$. All temperature values are in the units of $\varepsilon$. We use LAMMPS \cite{LAMMPS} to integrate the equations of motion with a timestep $0.005\tau$, while for the equilibration a larger time step of $0.01\tau$ was used. To improve the statistics we run $n=400$ independent copies of each system.

The mean-squared displacement of a chain was calculated as the time average
\begin{align}
\label{eq:g3_time_avg}
g(t,t_{\rm tot}) = \frac{1}{t_{\rm tot}-t} \int_{0}^{t_{\rm tot}-t} 
\left[ \mathbf{R}(t' + t) - \mathbf{R}(t') \right]^2 \textrm{d} t',
\end{align}
where $t_{\rm tot}$ is the total simulation length after discarding the initial $10^{5}\tau$ in which the steady state is reached. The vector $\mathbf{R}$ represents the position of the center of mass of the chain (or its hot/cold segment) for the $g_{3}$, or the position of a midpoint of the chain (or its hot/cold segment) for the $g_{1}$. The presented mean-squared displacements are an average over all $n$ chains.

\section{Supporting Information}
\label{sec:SI}
\begin{figure}[htb]
\includegraphics[height=5.5cm]{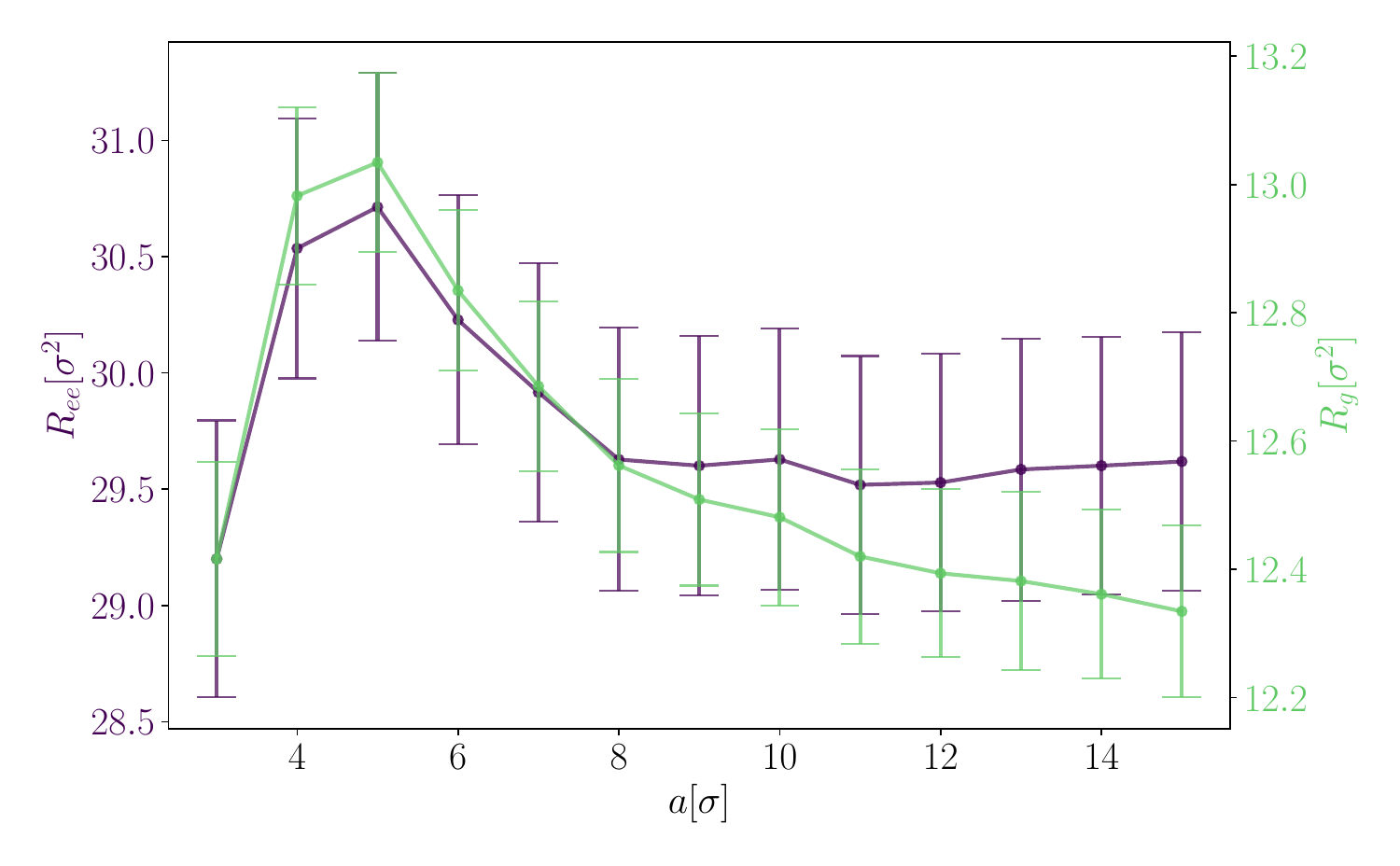}
\caption{\small Mean gyration radius $R_{g}$ and the end-to-end distance $R_{ee}$ of the systems as a function of the lattice spacing $a$. The mean performed over the steady state $t\geq 2.10^{5}\tau$. The error bars represent the standard deviation of the corresponding distribution. The ratio $R_{ee}^{2}/R_{g}^{2}\simeq 6$ as expected for a random walk.}
\label{fig:SI_Ree_Rg}
\centering
\end{figure}

\begin{figure}[htb]
\includegraphics[height=5.5cm]{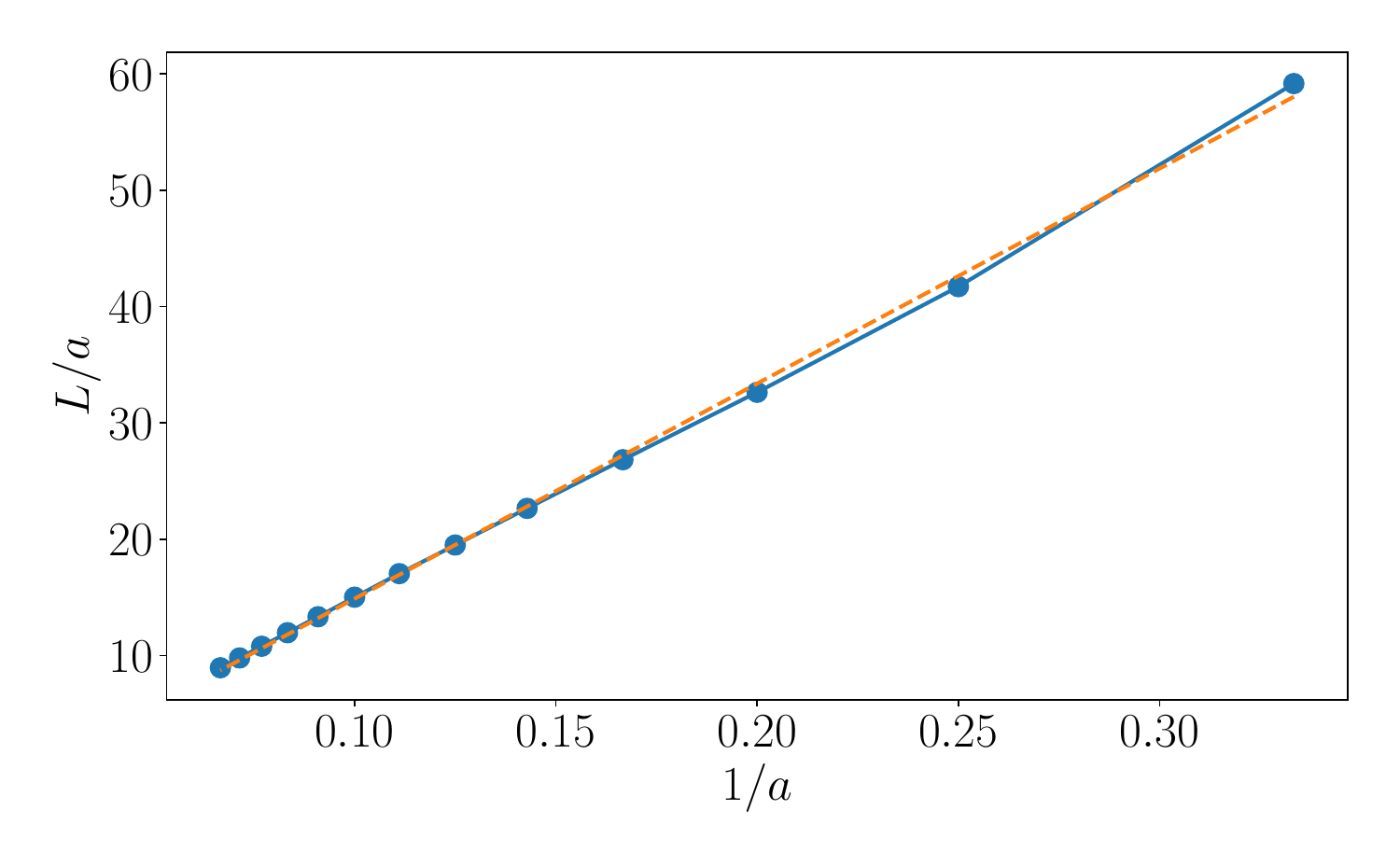}
\caption{\small The number of the lattice cells occupied by monomers of the chain without those involved in transverse excursion as a function of the inverse lattice spacing $1/a$. The quantity $L/a$ also represents the mean length of the primitive path/tube over $a$. The mean (blue symbols) is taken over all conformations at times $>10^{5}\tau$, spaced by $1.25\times 10^{4}\tau$. The error bars representing the error of the mean are smaller than the symbol size. The linear function (dashed orange) is a guide to the eye showing a linear relation ($L/a\sim a^{-1}$).}
\label{fig:SI_tube_length}
\centering
\end{figure}

\begin{figure}[htb]
\includegraphics[height=5.5cm]{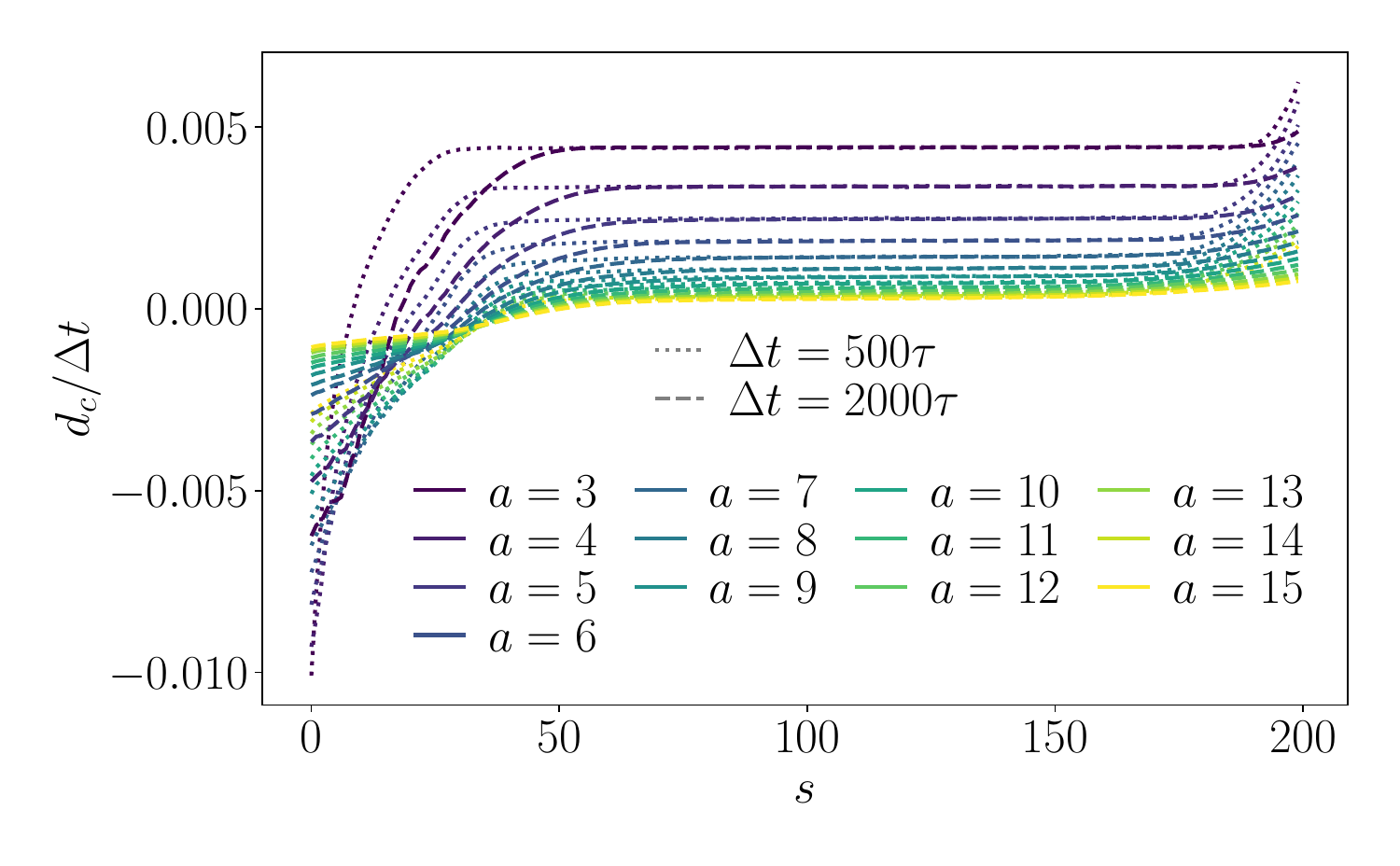}
\caption{\small Contour velocity $d_{c}/\Delta t$ as a function of contour coordinate $s$ for systems of different $a$ calculated from two different time lags $\Delta t = 500\tau$ (dotted) or $\Delta t = 2000\tau$ (dashed) for the systems with $T_{h}=3.0$.  The mean contour velocity is extracted as the mean over the region $s\in[75,125]$ where $d_{c}/\Delta t$ is constant and hence independent of the different time lags. The decrease (negative) contour displacements of the hot segment ($s<25$) and the upswing of the cold end are observation biases due to the end fluctuations. Only the fluctuating ends that do not enter the statistics of $v$ are affected by different $\Delta t$.}
\label{fig:SI_v_vs_Dt}
\centering
\end{figure}

\subsection{Drift/diffusion dominated relaxation and transverse displacements}
\label{sec:SI_coth}
\begin{figure}[htb]
\includegraphics[height=5.5cm]{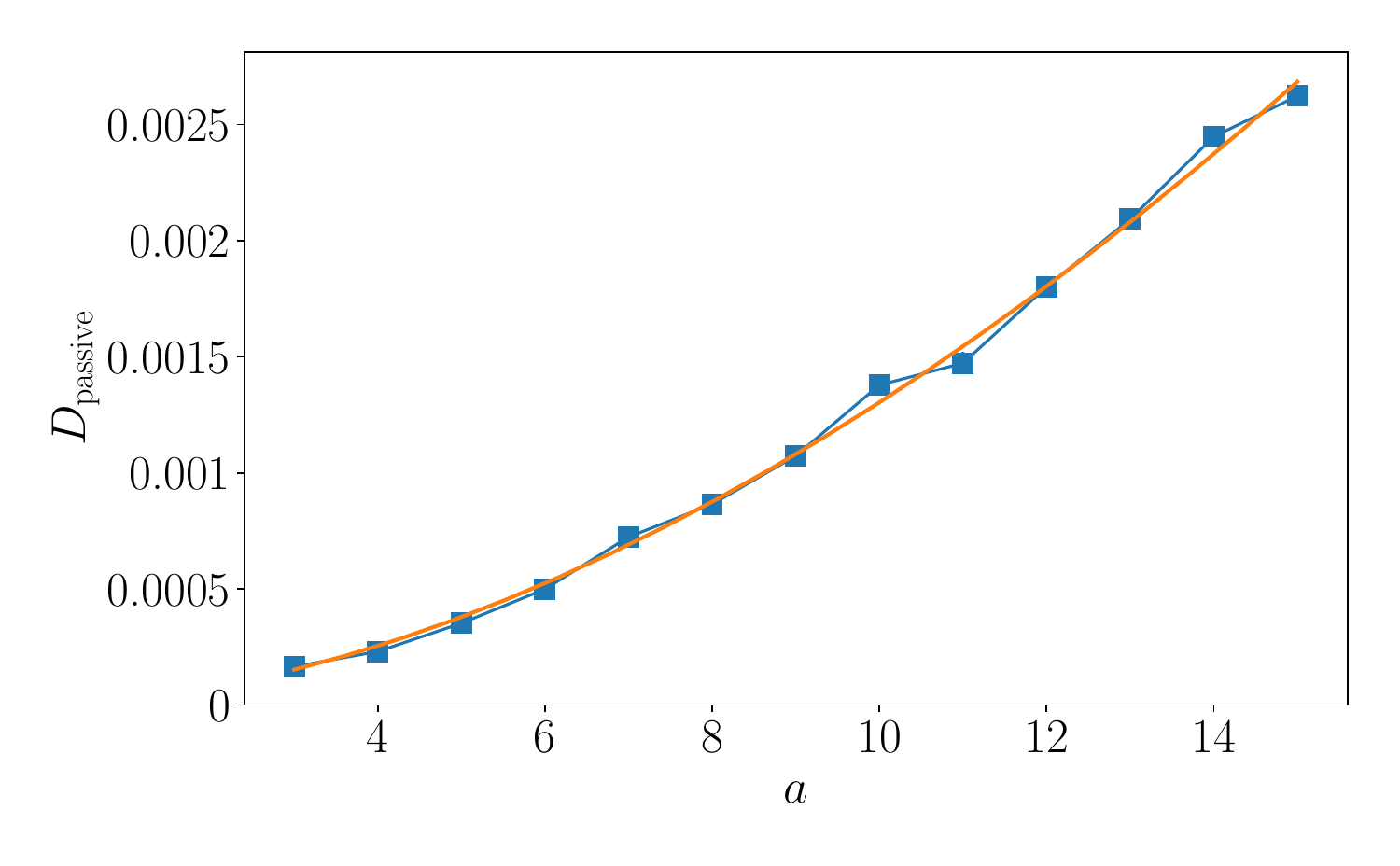}
\caption{\small Diffusion coefficient of the fully passive chains chains in equilibrium as a function of $a$ measured as $D = g_{3}/6t$ (symbols). Orange line represents the fitted power-law $D=2.155\times 10^{5} a^{1.78}$.}
\label{fig:SI_Deq}
\centering
\end{figure}

As mentioned in the main paper the diffusion coefficient, as calculated from the drift-diffusion equation \cite{Tejedor_Ramirez_SM20}, for a tangentially driven polymer is in a dimensionless form given by
\begin{align}
\label{eq:SI_DTej}
D = \frac{c}{6}\coth\frac{c Z}{2D_{c}},
\end{align}
where $Z$ is the number of lattice sites occupied by the chain. The dimensionless form is achieved by measuring the distances in lattice sites, and the time in the entanglement time. Hence the velocity $c$ is the number of lattice sites per entanglement time and the contour diffusion coefficient $D_{c}$ is the number of lattice sites squared per entanglement time. The argument of the $\coth$ corresponds to the ratio $\tau_{\rm diffusion}/\tau_{\rm drift}$, i.e. the time for the center of the chain to move out of the initial tube purely by drift $\tau_{\rm drift} = (Z/2)/c$ and by purely diffusion $\tau_{\rm diffusion} = (Z/2)^{2}/D_{c}$. In our units these times are $\tau_{\rm drift} = (N/2N_{e})/v$ and $\tau_{\rm diffusion} =  (a N/2N_{e})^{2}/D_{c}$, where $D_{c}$ has the units of $\sigma^{2}/\tau$ and $v$ just $\tau^{-1}$. Therefore $\tau_{\rm diffusion}/\tau_{\rm drift} = \frac{a^{2}N v}{2N_{e} D_{c}}$, which for our values of the parameters and $D_{c}\simeq 1/200$ gives $\tau_{\rm diffusion}/\tau_{\rm drift}$ in the range from about $50$ for $a=15$ to about $200$ for $a=3$, meaning that our systems are drift-dominated even for large $a$. This is confirmed by comparing the diffusion coefficient that we measured for the fully passive systems (Fig.~\ref{fig:SI_Deq}) to the ones with the activity Fig.~2 that are at least one order of magnitude larger.

\begin{figure}[htb]
\includegraphics[height=5.5cm]{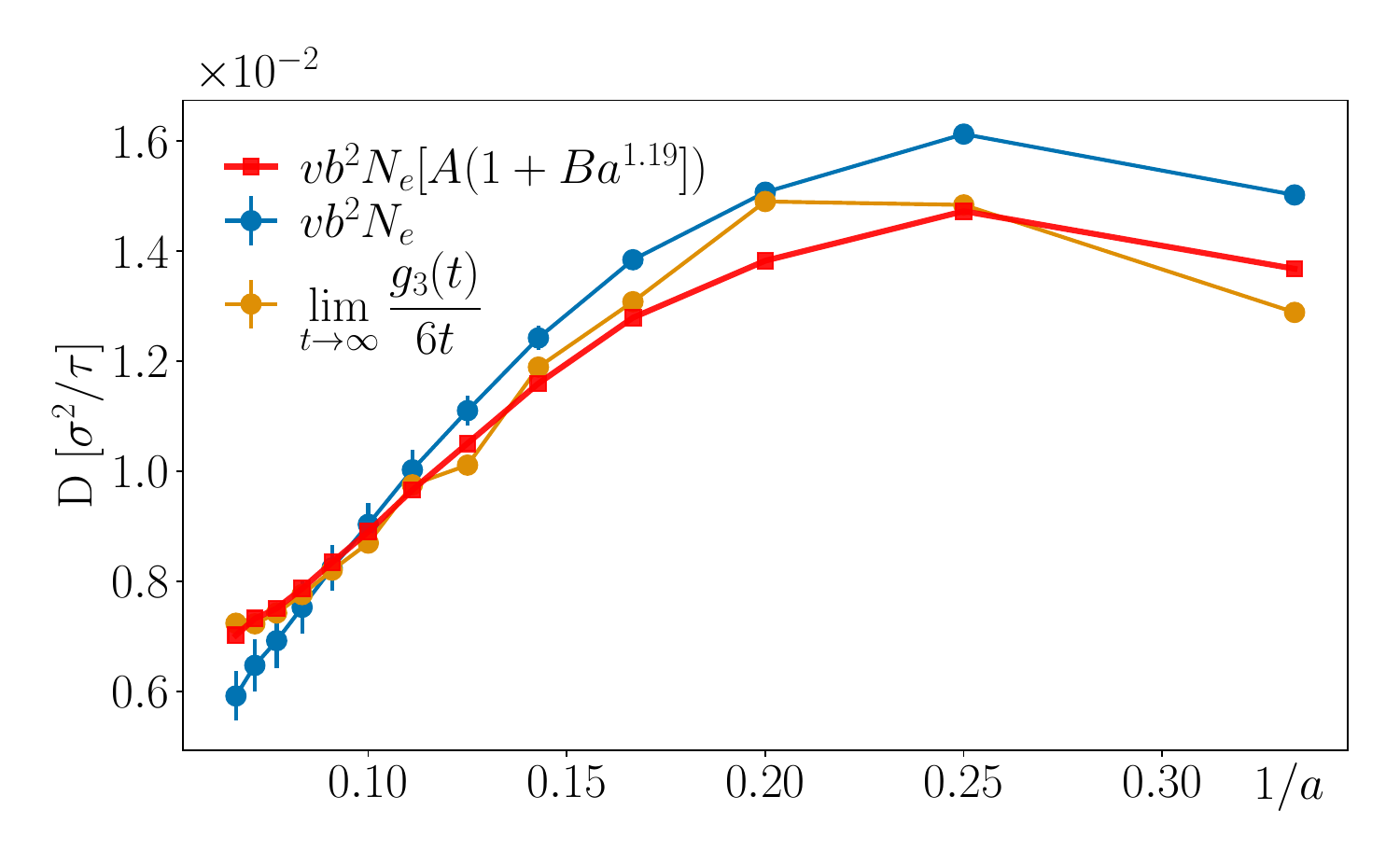}
\caption{\small Diffusion coefficient of the active system (orange), simple scaling estimate (blue) and the scaling estimate with the transverse displacement correction \eqref{eq:D_trans} (red) where we used the fitted form of $N_{e}\sim a^{1.19}$ from the main paper.}
\label{fig:SI_Dfit}
\centering
\end{figure}

In contrast to the tangentially driven chain on a lattice \cite{Tejedor_Ramirez_SM20}, our chains have the transverse freedom up to the cell size $a$. The transverse modes are responsible for the increased $D$ in comparison to the scaling estimate $v b^{2} N_{e}$ in cell sizes $a \geq 13$. Within the total relaxation time, the chain is relaxed on the scale and therefore can move about one cell size in the transverse direction. If in each cell the chain moves about squared distance $a^{2}$, and the displacements are independent, the center of mass squared displacement is about $a^{2}/(N/N_{e})$. The scaling estimate of the diffusion coefficient $D\sim R^{2}/\tau_{\rm relax}$ should be amended by such a term giving
\begin{align}
\label{eq:D_trans}
D \sim \frac{R^{2}+(a^{2}/(N/N_{e}))}{\tau_{\rm relax}}= v b^{2} N_{e}\left[1+\frac{a^{2}N_{e}}{b^{2}N^{2}}\right].
\end{align}
The term in the square bracket hence represents a correction factor to the original scaling estimate $v b^{2} N_{e}$. To see whether it can fit the data, we perform a two parameter fit $A[1+Ba^{2}N_{e}/b^{2}N^{2}]$ of $g_{3}/6t v b^{2} N_{e}$. The overall prefactor $A$ and the factor $B$ are used because the presented displacements are only scaling estimates and the displacements in different cells are not necessarily independent. As presented in Fig.~\ref{fig:SI_Dfit} the fit works relatively well in the regime of large $a$.

\begin{figure}[htb]
\includegraphics[height=5.5cm]{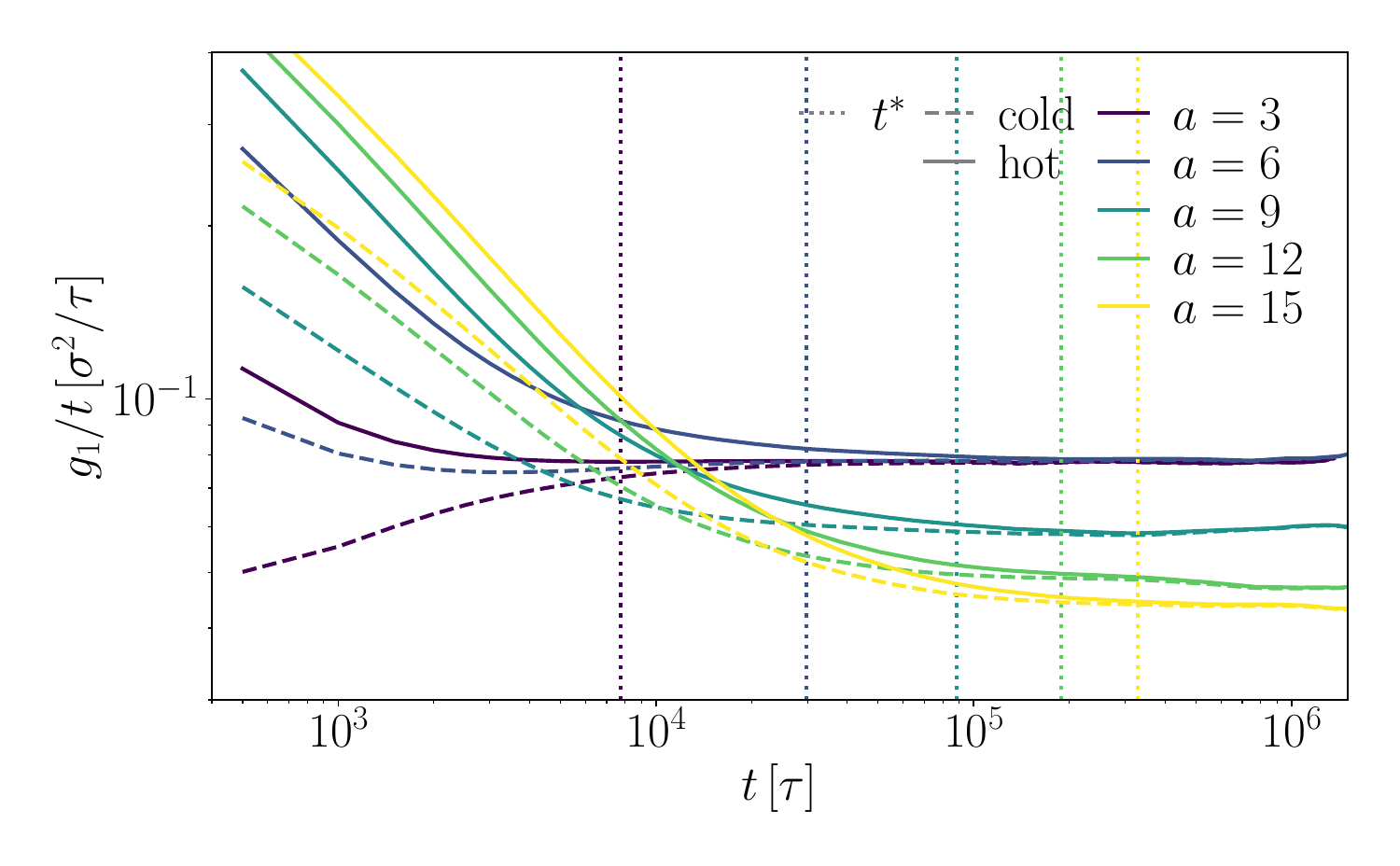}
\caption{\small Mean squared displacement over time $g_{1}/t$ of the \emph{first monomer} of the segment as a function of time. In contrast to the central monomer (see main text), the first monomer of the hot segment exhibits subdiffusive behavior at short times for all systems (even $a=3$) as expected.}
\label{fig:SI_g1_hot_head}
\centering
\end{figure}

\begin{figure*}[htb]
\includegraphics[height=5.5cm]{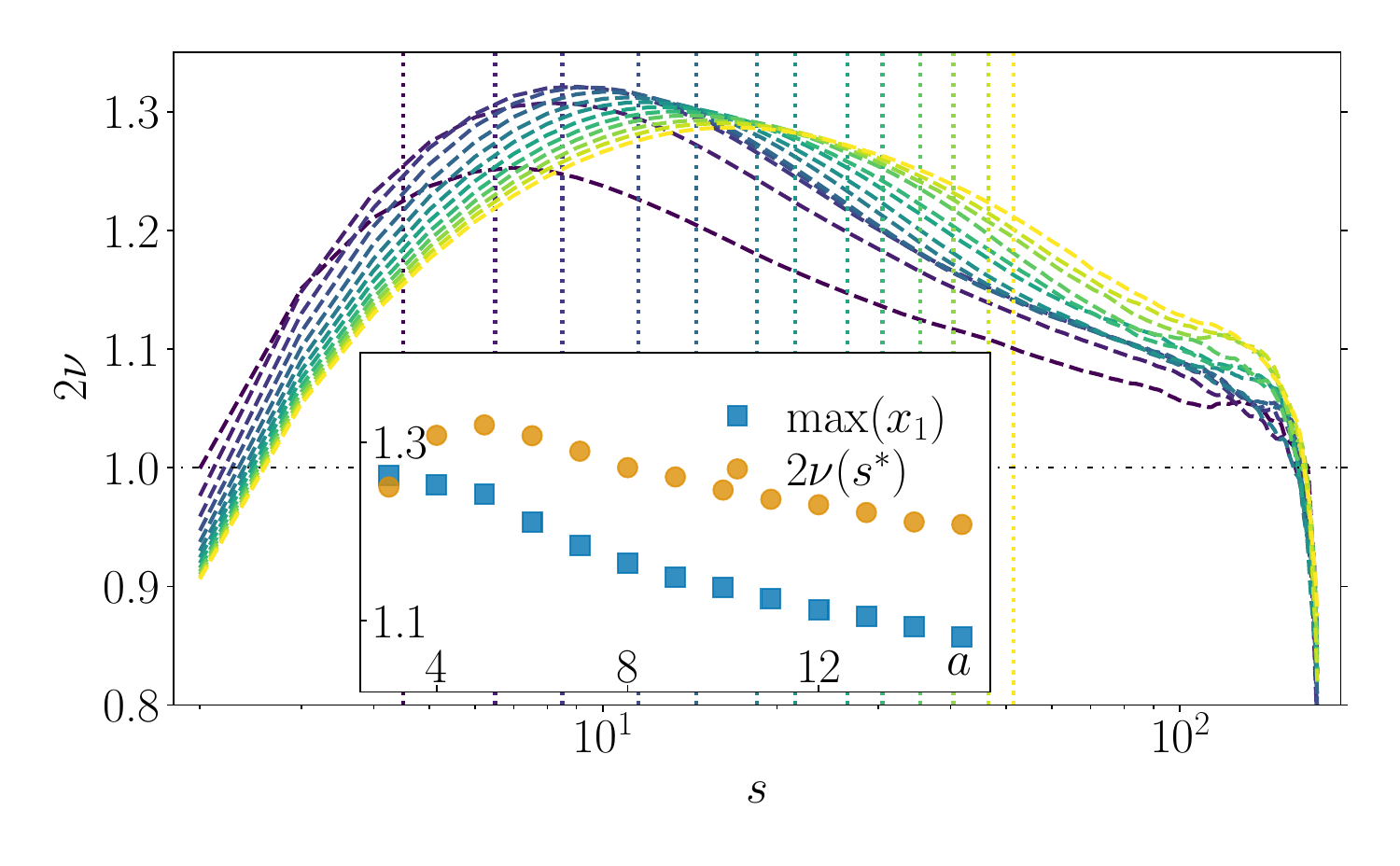}
\includegraphics[height=5.5cm]{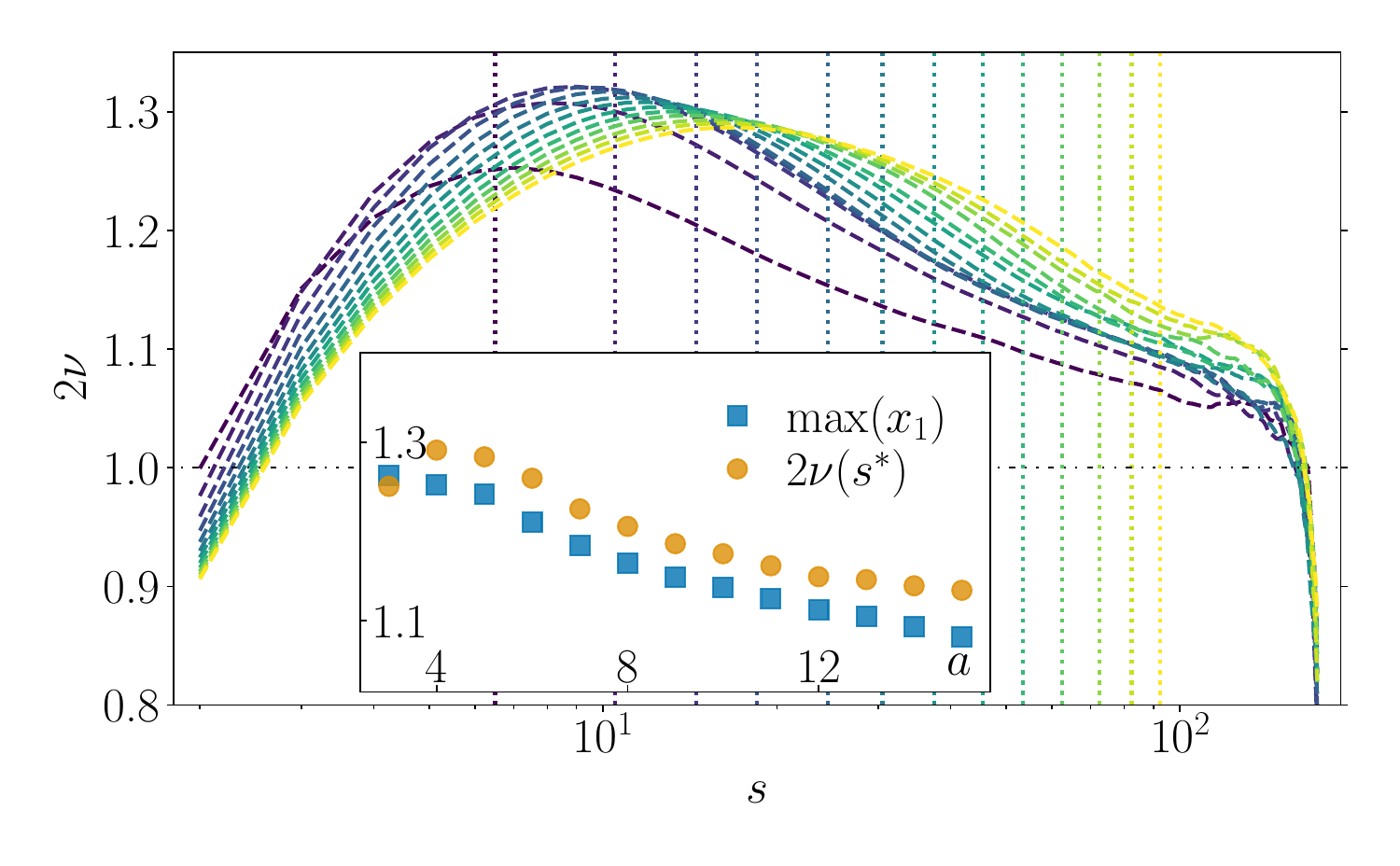}
\includegraphics[height=5.5cm]{x1max_vs_a.pdf}
\includegraphics[height=5.5cm]{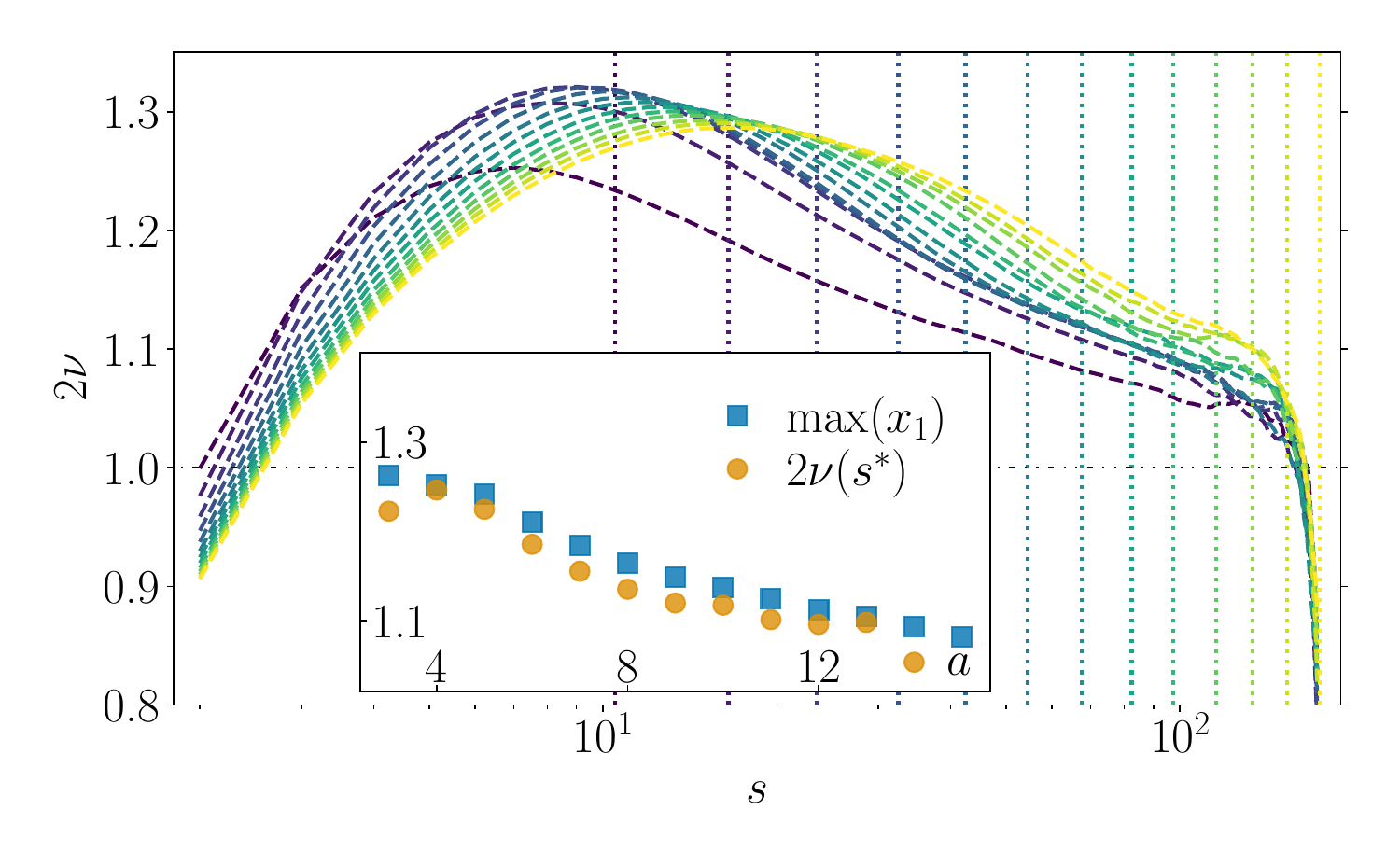}
\caption{\small The exponent $2\nu$ of the cold segment as a function of the contour distance $s$. The vertical dotted lines are at $s^{\ast}$ (see main text for details) extracted from when the means squared internal distance $d^{2}(s^{\ast}) = (k a)^{2}$. Here the top row corresponds to $k^{2}=1$ (left), $k^{2}=2$ (right), while the bottom row to $k^{2}=3$ (left - identical to the plot in the main paper) and $k^{2}=4$ (right). Insets: Comparison of the exponents $\max(x_{1})$ and $2\nu(s^{\ast})$. }
\label{fig:SI_x1_max_vs_a_vs_k}
\centering
\end{figure*}

\subsection{More details on $g_{3}$ calculation}
\label{sec:SI_g3}
\begin{figure}[h!tb]
\includegraphics[height=5.5cm]{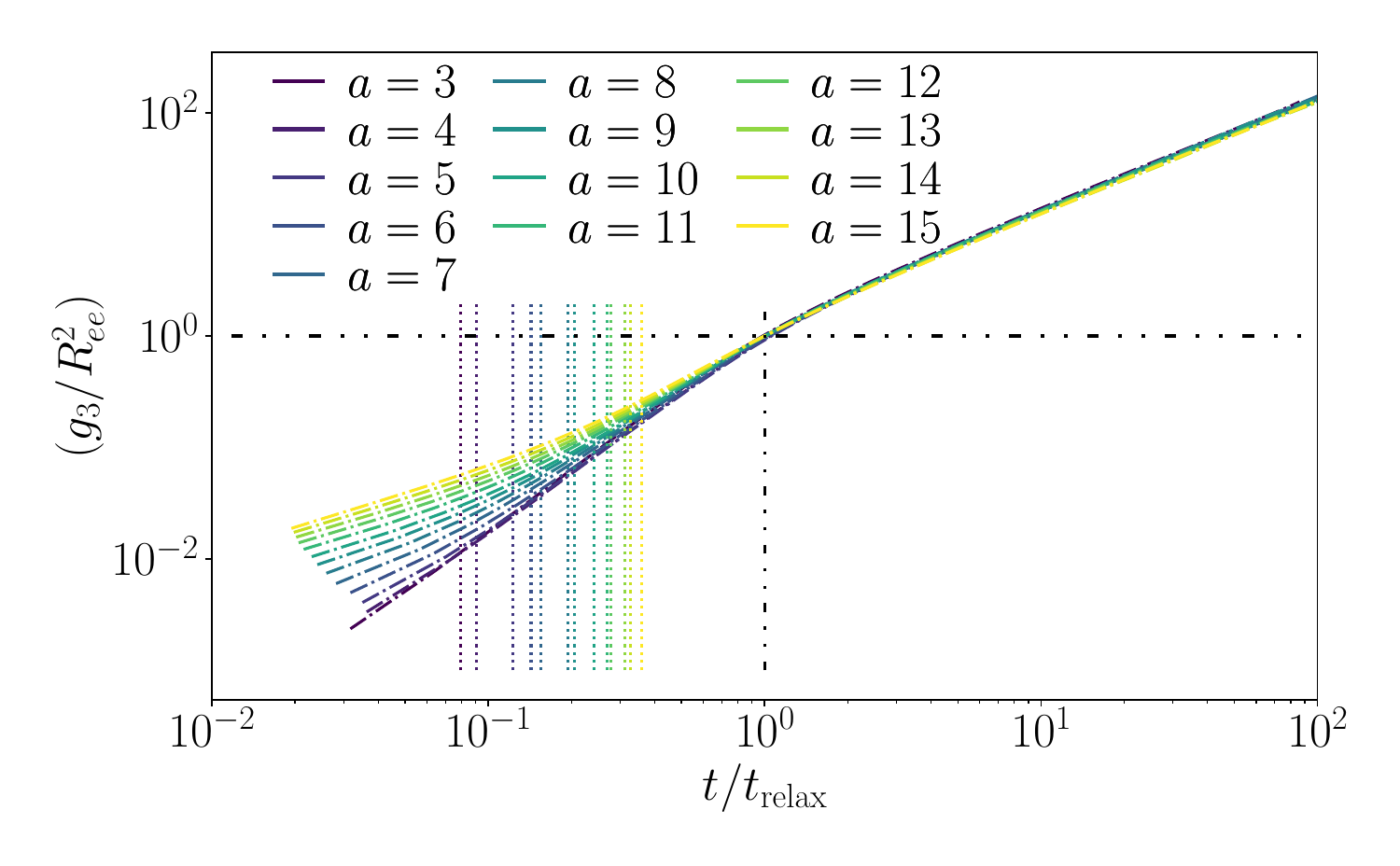}
\caption{\small Mean squared displacement of the chain's center of mass $g_{3}$ normalized by the mean squared end-to-end distance $R_{ee}^{2}$ as a function of time normalized by the time $t_{\rm relax}$, computed here as $g_{3}(t_{\rm relax}) = R_{ee}^{2}$. The vertical colored dotted lines show time when $g_{3}=a^{2}$, while the vertical black dash-dotted line is at $t/t_{\rm relax}=1$. The time span between the vertical colored and the black line coincide with superdiffusive regime of $g_{3}$. Longer superdiffusive regime is observed for systems with lower $a$.}
\label{fig:SI_g3_normalized_by_Ree}
\centering
\end{figure}

On time scales shorter than the relaxation time $t<\tau_{\rm relax}$, the lattice prevents rotational relaxation of the chain on the scale above the mesh size. Therefore, as the entropic pulling is at play, the mean squared displacement of the center of mass $g_{3}(t)$ is similar to that of the tangentially driven polymers \cite{Xiang_tangentially_driven_filaments_PRR23} exhibiting ballistic superdiffusion, but differs in the earlier crossover to diffusional regime. Below we recapitulate the calculation of \cite{Xiang_tangentially_driven_filaments_PRR23} and show where our system differs.

\begin{figure}[htb]
\includegraphics[width=0.45\textwidth]{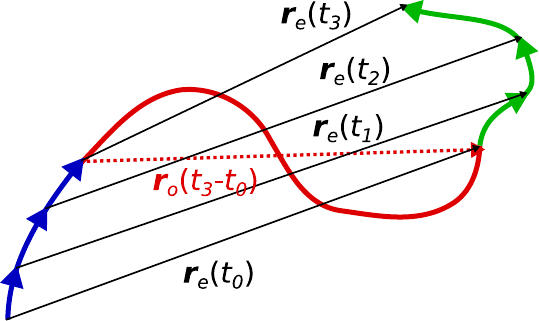}
\caption{\small The polymer path and its end-to-end vectors $ \bm{r}_{e}(t_{i})$ at different times $t_{i}$. The green part represents the path segment that the head of the polymer moved between $t_{0}$ and $t_{3}$, while the blue one similarly, but for the polymer tail. The red part represents the segments of the conformation that overlaps between $t_{0}$ and $t_{3}$ and the red dashed line is the corresponding end-to-end vector $\bm{r}_{o}$.}
\label{fig:SI_rcm}
\end{figure}

The center of mass position vector $\bm{r}_{\rm cm}(t) = N^{-1}\sum_{i=1}^{N} \bm{r}_{i}(t)$, with the monomer $i$ position vector $\bm{r}_{i}$. As the polymer moves along its contour path, the displacement $\Delta \bm{r}_{\rm cm}(t) = \bm{r}_{\rm cm}(t) -\bm{r}_{\rm cm}(0)$ can be expressed as the sum of the end-to-end vectors $\bm{r}_{e}$ of the whole polymer at times $t_{i}$ from $ t_{0}=0$ to $t_{k}=t$ (see Fig.~\ref{fig:SI_rcm})
\begin{align}
\Delta \bm{r}_{\rm cm}(t) = N^{-1}\sum_{i}^{k} \bm{r}_{e}(t_{i})
\end{align}
Switching to continuous representation $\sum_{i}^{k} \to \tau_{b}^{-1}\int_{0}^{t}dt^{\prime}$, where $\tau_{b}$ here is the unit of time (monomer to move its own size by drift: $b/v=\tau_{b}$), with $b$ the monomer size, the $g_{3}(t)$ is then
\begin{align}
\label{eq:g3a}
g_{3}(t) = (N\tau_{b})^{-2} \int_{0}^{t}\int_{0}^{t} \langle \bm{r}_{e}(t^{\prime})\bm{r}_{e}(t^{\prime\prime})\rangle dt^{\prime}dt^{\prime\prime}
\end{align}
At any time the end-to-end vector $\bm{r}_{e}(t_{i})$ can be decomposed in the sum ${\color{blue}\bm{b}}+{\color{red}\bm{r}}+{\color{green}\bm{g}}$, where the terms correspond to the sections of blue, red and green segments respectively. As the newly traced segments of the polymer path are uncorrelated to all the previous ones, i.e. $\langle {\color{green}\bm{g}}{\color{red}\bm{r}}\rangle = \langle {\color{green}\bm{g}}{\color{blue}\bm{b}}\rangle = \langle {\color{red}\bm{r}}{\color{blue}\bm{b}}\rangle = 0$ (Fig.~\ref{fig:SI_rcm}), the only contribution surviving the averaging in \eqref{eq:g3a} would be the overlapping (red) segment terms $\langle\bm{r}_{o}^{2}(t^{\prime\prime}-t^{\prime})\rangle = \bm{r}_{o}^{2}(t^{\prime\prime}-t^{\prime})$. Therefore, we can express $g_{3}$ as 
\begin{align}
\label{eq:g3b}
g_{3}(t) = (N\tau_{b})^{-2} \int_{0}^{t}\int_{0}^{t}  \bm{r}_{o}^{2}(t^{\prime\prime}-t^{\prime}) dt^{\prime}dt^{\prime\prime},
\end{align}
where the time dependence of the $\bm{r}^{2}_{o}(t)$ is 
\begin{align}
\label{eq:ro}
\bm{r}^{2}_{o}(t) = b^{2}\left[ N - \frac{v t}{b} \right].
\end{align}
Here $b$ is the bond length and we assumed ideal chain statistics and uniform driving velocity. Then, integrating \eqref{eq:g3b} gives 
the ballistic regime $g_{3}\sim t^{2}$ for $t< N\tau$ and the diffusive regime $g_{3}\sim t$ for $t > N\tau$.
 
\begin{figure}[h!]
\includegraphics[width=0.45\textwidth]{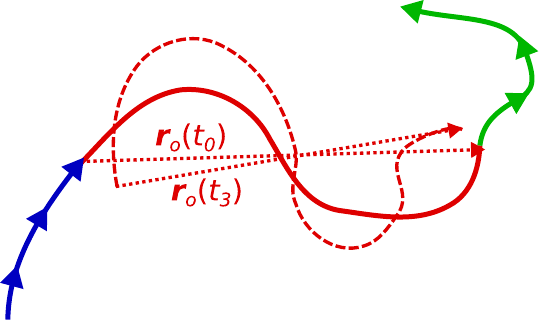}
\caption{\small The polymer between times $t_{0}$ and $t_{3}$ mostly follows the delineated path, but because of the transverse modes the overlapping (red) region is no longer perfectly correlated but can decorrelate.}
\label{fig:SI_rcm2}
\end{figure}

All the above holds because a tangentially driven chain follows closely its own path. In our case it holds true on the scales above $a$ because significant rotational diffusion is suppressed by the obstacles. However, when the overlap segment $\bm{r}_{o}$ spans just one cell size it can relax by rotational diffusion before it is pulled out completely. The criterion is $\tau_{a} > \tau_{e}$, that is $a/v > \tau_{0}N_{e}^{2}$, which for $v \simeq k_{B}\Delta T(a^{-1} - a_{\rm max}^{-1})/N\xi$ and $\tau_{0}=b^{2}k_{B}T_{c}/\xi$, gives the criterion in the form
\begin{align}
a^{2}\left(1-\frac{a}{a_{\rm max}}\right) < R^{2}\frac{T_{c}}{\Delta T},
\end{align}
where we used $R^{2}=Nb^{2}$. This inequality is satisfied for $a \ll a_{\rm max}$ (recall $a_{\rm max} \simeq R_{g}$) for not too strong temperature contrasts or when $a\simeq a_{\rm max}$. In either case, for the $\Delta T \simeq 2T_{c}$ that we use the chain is mostly relaxed on the scale of the cell. For small $a$, the chain spans many cells and therefore the crossover to ballistic regime can be observed, because the relaxation of $\bm{r}_{o}$ in the last cell does not affect the overall scaling. However, for large $a$ there are only few cells and the relaxation is relevant. As the transverse modes are significant at the chain end (see Fig.~\ref{fig:SI_rcm2}) the integrand of \eqref{eq:g3a} we approximate as $\langle \bm{r}_{e}(t^{\prime})\bm{r}_{e}(t^{\prime\prime})\rangle \simeq \langle  \bm{r}_{o}(t^{\prime\prime})\bm{r}_{o}(t^{\prime}) \rangle$ and further by asserting that
\begin{align}
  \langle  \bm{r}_{o}(t^{\prime\prime})\bm{r}_{o}(t^{\prime}) \rangle = \langle \bm{r}_{o}^{2}(t^{\prime\prime}-t^{\prime}) \rangle =  b^{2}\left[ N - \frac{v (t^{\prime\prime}-t^{\prime})}{b}\right]
\end{align}
only for $t^{\prime\prime}-t^{\prime} < t_{a}$ and $0$ otherwise, where $t_{a} = (N - N_{e})b/v$ is the time for the active force to pull out the polymer from its tube completely up to the last cell. In other words we assume that as the chain is completely relaxed below the tube the scalar product of $\bm{r}_{o}$ is zero in the last cell, but stays ideal in all other cells, because of the limited transverse relaxation above $a$. Then eq.~\eqref{eq:g3b} becomes
\begin{align}
\nonumber
g_{3}(t) = 2\frac{b^{2}}{N^{2}\tau^{2}} \int_{0}^{t-t_{a}} dt^{\prime} \int_{t^{\prime}}^{t_{a}+t}  [N-v(t^{\prime\prime}-t^{\prime})/b]dt^{\prime\prime} \\+ 2\frac{b^{2}}{N^{2}\tau^{2}} \int_{t-t_{a}}^{t} dt^{\prime} \int_{t^{\prime}}^{t}  [N-v(t^{\prime\prime}-t^{\prime})/b]dt^{\prime\prime}.
\end{align}
In the regime when $t_{a} < t < t_{\rm relax} = N\tau = Nb/v$, which is the regime when the chain is yet to be pulled out from the last cell, the result is 
\begin{align}
\label{eq:g3d}
g_{3}(t) =\frac{b^{2}}{N^{2}\tau^{2}} \left[  t_{a}(t - \frac{t_{a}}{2}) + \frac{v}{N b} \left( -t^{2}t_{a} + \frac{3}{2} t t_{a}^{2} - \frac{2}{3}t_{a}^{3} \right)\right].
\end{align}
In this regime the first term in \eqref{eq:g3d} dominates (their ratio is roughly $t_{\rm relax}/t$) and therefore $g_{3} \sim t$. The mean-squared displacement crosses over from superdiffusion to ordinary diffusion even before the whole tube is vacated due to Rouse relaxation of the chain end on the scale of $a$.

\bibliography{biblio}

\end{document}